\documentclass[twocolumn]{aastex701}

\usepackage{amsmath}
\usepackage{xcolor}
\shorttitle{Sample article}
\shortauthors{}
\graphicspath{{./}{Figures/}}

\begin{document}

\title{Compositions of Exoplanetary Debris Shaped by Milky Way Stellar Abundance Patterns}

\author[orcid=0000-0002-4872-1021]{Isabella L. Trierweiler}
\affiliation{Department of Astronomy, Yale University, New Haven, CT 06511, USA}
\email[show]{isabella.trierweiler@yale.edu}

\author[orcid=0000-0002-5937-9655]{Kendra Nguyen}
\affiliation{Department of Astronomy, Yale University, New Haven, CT 06511, USA}
\email[]{temp}

\author[orcid=0009-0002-6065-3292]{Lou~Baya~Ould~Rouis}
\affiliation{Department of Astronomy \& Institute for Astrophysical Research Boston University Boston, MA 02215, USA}
\email[]{temp}

\author[orcid=0000-0003-0965-605X]{Pratik Gandhi}
\affiliation{Department of Astronomy, Yale University, New Haven, CT 06511, USA}
\email[]{temp}

\author[orcid=0000-0002-3553-9474]{Laura K. Rogers}
\affiliation{NOIRLab, 950 N Cherry Ave., Tucson, AZ 85719, USA;}
\email[]{temp}

\author[orcid=0000-0002-9548-1526]{Carey M. Lisse}
\affiliation{Johns Hopkins University, 3400 N. Charles St, Baltimore, MD 21218, USA;}
\affiliation{Johns Hopkins University Applied Physics Laboratory, Laurel, MD 20723, USA}
\email[]{temp}

\author[orcid=0000-0002-7670-670X]{Malena Rice}
\affiliation{Department of Astronomy, Yale University, New Haven, CT 06511, USA}
\email[]{temp}

\author[0000-0002-5653-1543]{Damanveer S. Grewal}
\affiliation{Department of Earth and Planetary Sciences, Yale University, New Haven, CT 06511, USA}
\email[]{temp}

\author[0000-0002-1299-0801]{Edward Young}
\affiliation{Department of Earth, Planetary, and Space Sciences, University of California, Los Angeles, Los Angeles, CA 90095, USA}
\email[]{temp}


\begin{abstract}
Galactic chemical evolution (GCE) predicts large-scale variations in exoplanet compositions with host star elemental abundances, reflecting the Milky Way's history of star formation and elemental enrichment from stellar feedback. 
We investigate the impact of GCE on the compositions of exoplanetary bodies using polluted white dwarfs (WDs), which probe the bulk elemental abundances of planetary debris. 
Using a literature sample of 70 polluted WDs, we calculate iron mass fractions for the accreted bodies and compare these to tracers of the WD progenitor's initial abundances, including age and kinematics. 
We demonstrate predicted trends using a sample of main sequence stars, and we show that these are generally reflected in the polluted WD sample. 
In particular, H-dominated WDs show preferentially larger iron mass fractions amongst younger and less dynamically excited systems. 
This trend is weaker amongst He-dominated WDs, and we show that this may be attributed to uncertainties in their observed phases of accretion. 
Continuing to combine WD pollution compositions with stellar parameters and kinematics could be a valuable way to constrain large-scale versus local drivers of planetary compositions. 
\end{abstract}


\section{Introduction}\label{section:intro}

Exoplanet populations are expected to be strongly shaped by galactic chemical evolution \citep[GCE; see reviews by][]{Tinsley1980, Matteucci2021} as stellar nucleosynthesis and feedback processes shape the elemental abundances of host stars and the corresponding material available for planet formation \citep[e.g.,][]{Teske2024}. 
Dependencies between planet demographics and galactic context have been identified in metallicity-dependent occurrence rates \citep[e.g.,][]{Boley2024, Zink2023, Fischer2005} and in stellar elemental abundance patterns of exoplanet hosts \citep[e.g.,][]{Ghezzi2026, Sharma2024, Adibekyan2012}. 
At the same time, planetary compositions and structures  are predicted to correlate with stellar compositions \citep[e.g.,][]{Santos2017, Michel2020, Bitsch2020, Cabral2023, Behmard2025}.
For example, \citealt{Adibekyan2021} demonstrated that planetary iron-mass fractions, extrapolated from mass and radius measurements, correlate with the iron fractions of their stellar hosts. However, the certainty of this relationship remains debated \citep{Turner2026}, and if present, the exact nature of the star-planet composition connection can vary with interior structure of the planet \citep[e.g.][]{Chen2025}. 

Present-day planet compositions result from the convolution of both local planet formation processes and the long-term evolution of the raw material composition across the galaxy.
Insofar as exoplanet compositions reflect the elemental abundances of their host stars, the imprint of galactic evolution on planet demographics therefore encodes the cumulative effects of star formation, stellar feedback, and elemental enrichment of stars in the MW over cosmic time. 
Major mechanisms that shape GCE include O/B and AGB stellar winds, core-collapse and Type Ia supernovae, and a variety of r- and s-process production mechanisms such as neutron star mergers \citep{Sneden2008, Kobayashi2020}. 
One of the strongest features of GCE is a trend in the ratio of alpha-capture elements to iron [$\alpha$/Fe] relative to iron abundance [Fe/H] \citep[e.g.,][]{Anders2014, Hayden2015}. 
This trend is driven by the relative contributions of alpha-capture elements from AGB stars and core-collapse supernovae, and iron from Type Ia supernovae \citep{Weinberg2019, Kobayashi2020}.
Broadly, the stellar [$\alpha$/Fe] abundances in the Milky Way decrease over time as [Fe/H] increases, as early-time core-collapse supernova feedback gives way to later-time Type Ia supernovae feedback \citep{Tinsley1980, Kobayashi2020}.
This relation is especially relevant for exoplanet compositions as alpha-elements are typically lithophiles, partitioning into silicate or mantle material, while iron-peak elements correspond to siderophiles, which partition into metal cores. 
The change in stellar $\alpha$/Fe ratios therefore corresponds to long-term variation in the ratios of lithophile to siderophile elements available for star and planet formation and thus a hypothesized corresponding trend in core to mantle ratios for rocky exoplanets \citep[e.g.,][]{Michel2020, Trierweiler2023, Steffen2025, Weeks2025}. 

In this study, we focus on polluted white dwarfs (WDs) as tracers of the impact of GCE on exoplanetary compositions over time. 
Polluted WDs are stellar remnants which have accreted exoplanetary debris, enabling analyses of the bulk compositions of the accreted material \citep[e.g.,][]{Jura2014, Xu2024b}. 
Statistical analyses of the rock-forming elements polluting WDs suggests the accreted bodies typically have bulk abundance ratios similar to chondritic meteorites \citep{Doyle2023, Swan2023, Trierweiler2023}. 
We use the bulk element abundances measured for WD pollutants to carry out a test of iron mass fractions, as a proxy for core mass fractions (CMF), for the exoplanetary debris in polluted WD systems. 
In drawing this link between iron mass fraction and CMF, we assume that Fe is primarily partitioned into the metal cores of bodies, however we note that the amount of Fe stored in  cores can vary with the oxidation states at formation locations \citep{Doyle2019, Grewal2024}. 
Factors such as differentiation and thermal processing prior to accretion can also affect how strongly measured iron mass fractions correlate to parent body CMF \citep[e.g.,][]{Bonsor2020}. 


Using measured compositions of WD pollution from the literature, we map the bulk compositions of accreted planetary material to the iron mass fraction/core mass fractions (CMFs) of their parent bodies. 
We then pair these data with a tracer for the initial composition of the WD in order to roughly sort the WD systems by their ages and expected formation alpha/Fe ratio \citep[e.g.,][]{Raddi2022}. 
Lacking a direct means of measuring the initial [Fe/H] metallicity of the WD, we instead use (1) an estimate of the total age for the star based on the WD's current cooling age and inferred main sequence lifetime, and (2) a kinematic tracer of the WD's vertical orbital amplitude through the disk of the Milky Way, derived from \textit{Gaia} DR3 astrometry \citep{GaiaCollaboration2023}. 
Using these tracers, we test for a systematic variation in the CMF of accreted planetary debris with the progenitor properties of their WD hosts. 
We acknowledge that age determinations are notoriously difficult, and therefore use the two tracers independently without attempting to combine the results or weight one tracer over the other. 

In our literature sample of polluted WDs, we find evidence for trends between the CMF of pollution and the age/kinematics of the WD, with the strength of these trends varying between WDs with H- versus He-dominated atmospheres. 
We compare these results with expected trends based on a sample of main sequence stars drawn from the APOGEE/SDSS-V Milky Way Mapper DR 19 catalog \citep{Majewski2017, SDSSCollaboration2025}. 
Finally, we carry out a population synthesis model to estimate the polluted WD sample size required to reach a 3$\sigma$ difference in the CMFs of debris from old and young WD systems. 

This paper is organized as follows. In Section \ref{section:sample} we describe the assembly of the polluted WD and main sequence star samples used in this work. We describe the calculations for CMF and GCE tracers, as well as our methods for correlation testing, in Section \ref{section:methods} and present our results in Section \ref{section:results}. Finally, we discuss the impact of different WD atmospheric types, the possibility of preferential accretion of core versus mantle material, and uncertainties in pollution abundance measurements on our results in Section \ref{section:discussion}. 

\section{Sample of WDs and MS Stars}\label{section:sample}

\subsection{WD sample}
We include 70 polluted WD samples drawn from the literature and broadly group the WDs based on the dominant element in their atmospheres (H or He). 
Table \ref{table:WDsample} lists the identifiers and references for these objects. 
If an object had reported abundances from multiple sources, we generally selected the more recent publication or publication which reported the most characterized elements.
These systems are selected without placing any search limits on WDs parameters; we only require each sample to have measured Fe, Si, and Mg abundances in order to calculate iron/silicate mass ratios. 
Cross-checking our sample with the PEWDD \citep{Williams2024}, our sample includes all PEWDD systems to date with reported Fe, Si, and Mg abundances (omitting upper-limits). 

\startlongtable
\tabletypesize{\scriptsize}
\centerwidetable
\begin{deluxetable*}{l|l|l|l|l|l|l|l|l|l|l|l}
\tablecaption{WD sample; parameters are collected from the references listed in the table. Throughout this work we group WDs by the dominant element in their atmospheres (H or He), and we list them here in order of increasing $T_\mathrm{{eff}}$.}\label{table:WDsample}
\tablehead{\colhead{Name in Reference} & \colhead{Gaia DR3 Source ID} & \colhead{Type} & \colhead{$\mathrm{T_{eff}}$} & \colhead{log $g$} & \colhead{Reference\hyperlink{a}{$^a$}} & \colhead{CMF\hyperlink{b}{$^b$}} & \colhead{Age\hyperlink{c}{$^c$}} & \colhead{$v_{\mathrm{R}}$\hyperlink{d}{$^d$}} & \colhead{$\mathrm{R_{vel}}$\hyperlink{e}{$^e$}} & \colhead{$\mathrm{Z_{max}}$\hyperlink{f}{$^f$}} & \colhead{Min $\mathrm{Z_{max}}$\hyperlink{g}{$^g$}} \\*[-3mm]
\colhead{} & \colhead{} & \colhead{} & \colhead{(K)} & \colhead{(cm/s$^2$)} & \colhead{} & \colhead{$\%$} & \colhead{(Gyr)} & \colhead{(km/s)} & \colhead{(km/s)} & \colhead{(kpc)} & \colhead{(kpc)} }
\startdata
NLTT\,43806 & 1351956512512484480 & H & 5900$\pm$150 & 8.00$\pm${0.15} & \hyperlink{1}{1} & 13$\pm^{6}_{4}$ & 5.9$\pm^{4.5}_{1.4}$ &   &   &   & 0.02 \\
LSPM\,J0207+3331 & 325899163483416704 & H & 5910$\pm$98 & 8.11$\pm${0.03} & \hyperlink{2}{2} & 50$\pm^{10}_{10}$ & 4.8$\pm^{1.7}_{0.6}$ &   &   &   & 0.02 \\
WD\,2157-574 & 6412251624488840448 & H & 7010$\pm$250 & 8.06$\pm${0.08} & \hyperlink{3}{3} & 29$\pm^{6}_{5}$ & 4.6$\pm^{4.9}_{1.5}$ &   &   &   & 0.02 \\
WD\,2115-560 & 6462911897617050240 & H & 9600$\pm$340 & 7.97$\pm${0.08} & \hyperlink{3}{3} & 37$\pm^{7}_{7}$ & 5.7$\pm^{5.1}_{3.1}$ &   &   &   & 0.02 \\
G29$-$38 & 2660358032257156736 & H & 11820$\pm$200 & 8.40$\pm${0.10} & \hyperlink{4}{4} & 30$\pm^{8}_{7}$ & 1.0$\pm^{0.1}_{0.1}$ & -18.4 & 43.9$\pm^{4.0}_{4.0}$ & 0.31$\pm^{0.14}_{0.13}$ & 0.01 \\
WD\,0145+234 & 291057843317534464 & H & 12680$\pm$80 & 8.06$\pm${0.01} & \hyperlink{5}{5} & 45$\pm^{15}_{14}$ & 2.2$\pm^{2.2}_{0.6}$ & 10.5 & 3.6$\pm^{24.4}_{24.4}$ & 0.13$\pm^{0.10}_{0.09}$ & 0.02 \\
GD\,56 & 3251748915515143296 & H & 15270$\pm$300 & 8.09$\pm${0.10} & \hyperlink{6}{6} & 59$\pm^{14}_{15}$ & 2.2$\pm^{4.5}_{1.1}$ & -15.3 & 41.8$\pm^{8.9}_{8.9}$ & 0.07$\pm^{0.04}_{0.03}$ & 0.04 \\
WD\,0842+572 & 1037258898615762048 & H & 15400$\pm$70 & 7.91$\pm${0.01} & \hyperlink{5}{5} & 29$\pm^{13}_{10}$ & 7.8$\pm^{3.8}_{3.5}$ & -0.9 & 13.7$\pm^{4.7}_{4.7}$ & 0.10$\pm^{0.03}_{0.02}$ & 0.08 \\
J0611$-$6931 & 5279484614703730944 & H & 16530$\pm$561 & 7.81$\pm${0.03} & \hyperlink{7}{7} & 23$\pm^{6}_{5}$ &   & 36.6 & 56.8$\pm^{7.2}_{7.2}$ & 0.09$\pm^{0.06}_{0.03}$ & 0.06 \\
HE\,0106$-$3253 & 5026963661794939520 & H & 17350$\pm$200 & 8.12$\pm${0.10} & \hyperlink{8}{8} & 88$\pm^{2}_{2}$ & 1.6$\pm^{3.6}_{0.8}$ & 19.3 & 7.4$\pm^{15.3}_{15.3}$ & 0.16$\pm^{0.15}_{0.08}$ & 0.07 \\
SDSS\,J1043+0855 & 3869060540584643328 & H & 18330$\pm$340 & 8.05$\pm${0.06} & \hyperlink{9}{9} & 12$\pm^{13}_{7}$ & 2.1$\pm^{3.7}_{0.9}$ & 6.2 & 49.8$\pm^{4.9}_{4.9}$ & 0.28$\pm^{0.12}_{0.10}$ & 0.14 \\
PG\,1015+161 & 3888723386196630784 & H & 19200$\pm$180 & 8.22$\pm${0.06} & \hyperlink{10}{10} & 65$\pm^{15}_{19}$ & 0.7$\pm^{0.4}_{0.1}$ & 25.2 & 41.0$\pm^{5.0}_{5.0}$ & 0.10$\pm^{0.09}_{0.03}$ & 0.06 \\
WD\,1943+163 & 1820678800528443776 & H & 19250$\pm$30 & 7.87$\pm${0.01} & \hyperlink{11}{11} & 42$\pm^{12}_{11}$ & 8.4$\pm^{3.5}_{3.6}$ & 5.7 & 59.8$\pm^{7.4}_{7.4}$ & 0.27$\pm^{0.02}_{0.02}$ & 0.00 \\
WD\,1953–715 & 6422236358302245248 & H & 19250$\pm$60 & 8.12$\pm${0.02} & \hyperlink{11}{11} & 50$\pm^{13}_{14}$ & 1.0$\pm^{0.5}_{0.3}$ & -18.5 & 44.1$\pm^{7.3}_{7.3}$ & 0.28$\pm^{0.09}_{0.08}$ & 0.03 \\
WD\,0059+257 & 307372225052233984 & H & 20110$\pm$40 & 7.89$\pm${0.02} & \hyperlink{11}{11} & 79$\pm^{7}_{9}$ & 7.5$\pm^{4.1}_{3.5}$ & 19.4 & 25.8$\pm^{9.1}_{9.1}$ & 0.58$\pm^{0.15}_{0.13}$ & 0.06 \\
SDSS\,1228+1040 & 3904415787947492096 & H & 20900$\pm$900 & 8.15$\pm${0.04} & \hyperlink{10}{10} & 50$\pm^{19}_{18}$ & 0.9$\pm^{0.6}_{0.2}$ & -0.1 & 23.3$\pm^{2.9}_{2.9}$ & 0.15$\pm^{0.05}_{0.03}$ & 0.12 \\
GALEX\,1931+0117 & 4287654959563143168 & H & 21200$\pm$50 & 7.91$\pm${0.02} & \hyperlink{10}{10}, \hyperlink{12}{12} & 47$\pm^{17}_{18}$ & 6.5$\pm^{4.4}_{3.3}$ &   &   &   & 0.01 \\
J2100+2122 & 1837948790953103232 & H & 22000$\pm$399 & 7.92$\pm${0.02} & \hyperlink{7}{7} & 47$\pm^{9}_{9}$ & 5.9$\pm^{4.6}_{3.1}$ & -22.1 & 13.9$\pm^{7.9}_{7.9}$ & 0.04$\pm^{0.02}_{0.01}$ & 0.02 \\
WD\,1647+375 & 1351486711809893888 & H & 22040$\pm$31 & 7.88$\pm${0.03} & \hyperlink{13}{13} & 33$\pm^{11}_{10}$ & 7.2$\pm^{4.2}_{3.5}$ & -10.8 & 14.4$\pm^{10.2}_{10.2}$ & 0.27$\pm^{0.12}_{0.10}$ & 0.05 \\
PG\,0843+516 & 1029081452683108480 & H & 23095$\pm$230 & 8.17$\pm${0.06} & \hyperlink{10}{10} & 76$\pm^{9}_{12}$ & 0.7$\pm^{0.6}_{0.2}$ & 41.1 & 58.4$\pm^{5.9}_{5.9}$ & 0.47$\pm^{0.14}_{0.12}$ & 0.08 \\
\hline
SDSS\,J1535+1247 & 1193520666521113344 & He & 5680$\pm$50 & 8.06$\pm${0.03} & \hyperlink{14}{14} & 5$\pm^{4}_{2}$ & 7.9$\pm^{3.7}_{1.9}$ &   &   &   & 0.01 \\
SDSS\,J1038$-$0036 & 3806714593452089856 & He & 7560$\pm$150 & 8.06$\pm${0.04} & \hyperlink{14}{14} & 6$\pm^{2}_{1}$ & 5.2$\pm^{4.8}_{2.1}$ &   &   &   & 0.05 \\
J0956 & 1049598202217308160 & He & 8720$\pm$100 & 8.13$\pm${0.05} & \hyperlink{15}{15} & 5$\pm^{3}_{2}$ & 2.8$\pm^{2.8}_{0.7}$ &   &   &   & 0.10 \\
1252+7352 & 1691333347793733632 & He & 8785$\pm$110 & 8.17$\pm${0.02} & \hyperlink{16}{16} & 22$\pm^{9}_{7}$ & 2.2$\pm^{0.5}_{0.4}$ &   &   &   & 0.11 \\
1352+0323 & 3665500703030079488 & He & 9615$\pm$100 & 7.93$\pm${0.07} & \hyperlink{16}{16} & 15$\pm^{1}_{1}$ & 7.6$\pm^{4.4}_{3.7}$ &   &   &   & 0.23 \\
WD\,0446$-$255 & 4881758307940843008 & He & 10120$\pm$200 & 8.00$\pm${0.04} & \hyperlink{3}{3} & 19$\pm^{5}_{4}$ & 6.5$\pm^{4.7}_{3.2}$ &   &   &   & 0.06 \\
J2050 & 4226851932184135552 & He & 10200$\pm$300 & 8.44$\pm${0.21} & \hyperlink{15}{15} & 6$\pm^{7}_{3}$ & 1.8$\pm^{1.4}_{0.3}$ &   &   &   & 0.13 \\
0452$-$0214 & 3226269756470965120 & He & 10470$\pm$170 & 7.92$\pm${0.03} & \hyperlink{16}{16} & 13$\pm^{1}_{1}$ & 9.2$\pm^{3.4}_{3.7}$ &   &   &   & 0.07 \\
GD\,362 & 1336442472164656000 & He & 10540$\pm$200 & 8.24$\pm${0.04} & \hyperlink{17}{17}, \hyperlink{18}{18} & 48$\pm^{13}_{13}$ & 1.4$\pm^{0.4}_{0.2}$ & 5.3 & 55.6$\pm^{5.2}_{5.2}$ & 0.08$\pm^{0.07}_{0.04}$ & 0.03 \\
PG\,1225$-$079 & 3583181371265430656 & He & 10800$\pm$300 & 8.00$\pm${0.30} & \hyperlink{19}{19}, \hyperlink{18}{18} & 36$\pm^{7}_{6}$ & 3.7$\pm^{6.0}_{2.1}$ & 16.2 & 17.0$\pm^{9.0}_{9.0}$ & 0.22$\pm^{0.13}_{0.12}$ & 0.03 \\
WD\,1350$-$162 & 6295510766956001024 & He & 11640$\pm$290 & 8.02$\pm${0.05} & \hyperlink{3}{3} & 32$\pm^{7}_{7}$ & 5.3$\pm^{5.0}_{2.8}$ &   &   &   & 0.08 \\
WD\,1232+563 & 1571584539980588544 & He & 11787$\pm$423 & 8.30$\pm${0.06} & \hyperlink{8}{8} & 25$\pm^{6}_{5}$ & 1.1$\pm^{0.2}_{0.1}$ & -28.7 & 42.7$\pm^{6.0}_{6.0}$ & 0.35$\pm^{0.15}_{0.12}$ & 0.14 \\
1626+3136 & 1324451679388364288 & He & 12200$\pm$230 & 8.14$\pm${0.07} & \hyperlink{16}{16} & 22$\pm^{1}_{1}$ & 2.0$\pm^{3.3}_{0.7}$ &   &   &   & 0.22 \\
0242+0426 & 5743642585443328 & He & 12290$\pm$285 & 7.99$\pm${0.06} & \hyperlink{16}{16} & 14$\pm^{2}_{1}$ & 5.9$\pm^{5.0}_{3.2}$ &   &   &   & 0.18 \\
0850+3208 & 706974637946866304 & He & 12596$\pm$120 & 8.15$\pm${0.02} & \hyperlink{16}{16} & 41$\pm^{2}_{2}$ & 1.6$\pm^{0.6}_{0.4}$ &   &   &   & 0.07 \\
1333+3254 & 1469124246843402624 & He & 12830$\pm$50 & 7.99$\pm${0.01} & \hyperlink{16}{16} & 18$\pm^{2}_{2}$ & 6.7$\pm^{4.4}_{3.2}$ &   &   &   & 0.31 \\
SDSS\,J2339$-$0424 & 2446993162322393088 & He & 13735$\pm$500 & 7.93$\pm${0.09} & \hyperlink{21}{21} & 18$\pm^{4}_{4}$ & 6.3$\pm^{4.8}_{3.6}$ & 6.1 & 6.4$\pm^{10.9}_{10.9}$ & 0.12$\pm^{0.11}_{0.04}$ & 0.08 \\
SDSS\,J0738+1835 & 671450448046315520 & He & 13950$\pm$100 & 8.40$\pm${0.20} & \hyperlink{22}{22} & 26$\pm^{6}_{5}$ & 1.0$\pm^{2.3}_{0.2}$ &   &   &   & 0.06 \\
HS\,2253+8023 & 2286107295188538240 & He & 14400$\pm$400 & 8.40$\pm${0.30} & \hyperlink{19}{19} & 35$\pm^{3}_{3}$ & 1.4$\pm^{4.6}_{0.6}$ & -51.9 & 46.2$\pm^{7.4}_{7.4}$ & 0.17$\pm^{0.04}_{0.04}$ & 0.02 \\
WD\,1425+540 & 1608497864040134016 & He & 14490$\pm$345 & 7.95$\pm${0.08} & \hyperlink{23}{23} & 31$\pm^{13}_{11}$ & 6.1$\pm^{4.8}_{3.5}$ & -24.1 & 69.5$\pm^{3.5}_{3.5}$ & 0.09$\pm^{0.09}_{0.04}$ & 0.04 \\
0944$-$0039 & 3827999107046766720 & He & 14607$\pm$45 & 8.76$\pm${0.02} & \hyperlink{24}{24} & 28$\pm^{4}_{4}$ & 0.9$\pm^{0.0}_{0.0}$ &   &   &   & 0.10 \\
Gaia\,J0218+3625 & 328152307624540032 & He & 14700$\pm$500 & 7.86$\pm${0.05} & \hyperlink{25}{25} & 22$\pm^{8}_{7}$ &   & 16.1 & 18.0$\pm^{7.3}_{7.3}$ & 0.44$\pm^{0.09}_{0.08}$ & 0.04 \\
EC\,22211$-$2525 & 6625468025993283200 & He & 14740$\pm$500 & 7.89$\pm${0.05} & \hyperlink{25}{25} & 32$\pm^{8}_{7}$ & 8.4$\pm^{3.7}_{3.8}$ & 22.5 & 12.2$\pm^{13.3}_{13.3}$ & 0.27$\pm^{0.15}_{0.12}$ & 0.10 \\
WD\,2207+121 & 2727904257071365760 & He & 14752$\pm$1192 & 7.97$\pm${0.12} & \hyperlink{8}{8} & 22$\pm^{6}_{5}$ & 5.0$\pm^{5.4}_{3.1}$ & 6.4 & 20.3$\pm^{5.0}_{5.0}$ & 0.13$\pm^{0.06}_{0.03}$ & 0.10 \\
WD\,1551+175 & 1196531988354226560 & He & 14756$\pm$1286 & 8.02$\pm${0.12} & \hyperlink{8}{8} & 22$\pm^{5}_{4}$ & 4.1$\pm^{5.6}_{2.5}$ & -7.6 & 32.4$\pm^{9.0}_{9.0}$ & 0.40$\pm^{0.14}_{0.12}$ & 0.13 \\
WD\,1244+498 & 1567391625404274304 & He & 15150$\pm$500 & 7.97$\pm${0.05} & \hyperlink{25}{25} & 51$\pm^{10}_{9}$ & 6.3$\pm^{4.7}_{3.4}$ & 11.0 & 31.6$\pm^{3.9}_{3.9}$ & 0.19$\pm^{0.15}_{0.06}$ & 0.13 \\
WD\,1248+1004 & 3735027667976597248 & He & 15180$\pm$500 & 8.11$\pm${0.05} & \hyperlink{25}{25} & 30$\pm^{9}_{8}$ & 2.0$\pm^{3.0}_{0.7}$ & 5.7 & 17.4$\pm^{4.6}_{4.6}$ & 0.19$\pm^{0.11}_{0.03}$ & 0.16 \\
G241$-$6 & 2225982838287401984 & He & 15300$\pm$500 & 8.00$\pm${0.30} & \hyperlink{26}{26} & 18$\pm^{7}_{5}$ & 3.3$\pm^{6.0}_{2.2}$ & -55.0 & 96.3$\pm^{8.3}_{8.3}$ & 0.67$\pm^{0.06}_{0.04}$ & 0.10 \\
GD\,40 & 5187830356195791488 & He & 15300$\pm$500 & 8.00$\pm${0.30} & \hyperlink{26}{26} & 27$\pm^{10}_{9}$ & 3.3$\pm^{6.0}_{2.2}$ & -15.0 & 34.3$\pm^{8.5}_{8.5}$ & 0.10$\pm^{0.05}_{0.04}$ & 0.05 \\
1516$-$0040 & 4418628372344562048 & He & 15448$\pm$20 & 8.42$\pm${0.01} & \hyperlink{24}{24} & 32$\pm^{4}_{4}$ & 0.7$\pm^{0.0}_{0.0}$ &   &   &   & 0.10 \\
Gaia\,J1922+4709 & 2127665711125011456 & He & 15500$\pm$500 & 7.95$\pm${0.05} & \hyperlink{25}{25} & 46$\pm^{11}_{10}$ & 6.9$\pm^{4.5}_{3.6}$ & 17.6 & 32.7$\pm^{6.0}_{6.0}$ & 0.15$\pm^{0.05}_{0.04}$ & 0.03 \\
WD\,1145+017 & 3796414192429498880 & He & 15500$\pm$500 & 8.19$\pm${0.03} & \hyperlink{27}{27} & 56$\pm^{17}_{20}$ & 1.1$\pm^{0.5}_{0.3}$ &   &   &   & 0.13 \\
GD\,378 & 2109852523240471936 & He & 15620$\pm$500 & 7.93$\pm${0.06} & \hyperlink{21}{21} & 33$\pm^{4}_{4}$ & 7.0$\pm^{4.4}_{3.7}$ & -0.6 & 14.1$\pm^{9.2}_{9.2}$ & 0.23$\pm^{0.08}_{0.07}$ & 0.02 \\
0859+1123 & 603986308646058496 & He & 15717$\pm$63 & 8.19$\pm${0.04} & \hyperlink{24}{24} & 10$\pm^{4}_{3}$ & 1.1$\pm^{0.5}_{0.3}$ &   &   &   & 0.18 \\
0030+1526 & 2780715106223331072 & He & 15795$\pm$27 & 8.18$\pm${0.02} & \hyperlink{24}{24} & 23$\pm^{9}_{7}$ & 1.1$\pm^{0.5}_{0.3}$ &   &   &   & 0.11 \\
0930+0618 & 3852675309069466496 & He & 15982$\pm$41 & 8.18$\pm${0.02} & \hyperlink{24}{24} & 20$\pm^{4}_{3}$ & 1.1$\pm^{0.5}_{0.3}$ &   &   &   & 0.14 \\
1627+1723 & 4466719602193586816 & He & 16134$\pm$102 & 8.29$\pm${0.05} & \hyperlink{24}{24} & 44$\pm^{15}_{15}$ & 0.7$\pm^{0.2}_{0.1}$ &   &   &   & 0.16 \\
1109+1318 & 3965233688795064832 & He & 16308$\pm$62 & 8.25$\pm${0.03} & \hyperlink{24}{24} & 34$\pm^{9}_{8}$ & 0.8$\pm^{0.3}_{0.1}$ &   &   &   & 0.23 \\
SDSS\,J1734+6052 & 1435966347699773952 & He & 16340$\pm$500 & 8.04$\pm${0.05} & \hyperlink{25}{25} & 30$\pm^{10}_{8}$ & 3.8$\pm^{4.8}_{2.1}$ & -13.9 & 21.0$\pm^{8.7}_{8.7}$ & 0.28$\pm^{0.10}_{0.09}$ & 0.07 \\
0259$-$0721 & 5179698505635671680 & He & 16390$\pm$28 & 8.26$\pm${0.02} & \hyperlink{24}{24} & 13$\pm^{4}_{3}$ & 0.7$\pm^{0.2}_{0.1}$ &   &   &   & 0.15 \\
GD\,424 & 547179765520150912 & He & 16560$\pm$75 & 8.25$\pm${0.02} & \hyperlink{28}{28} & 22$\pm^{4}_{4}$ & 0.7$\pm^{0.2}_{0.1}$ &   &   &   & 0.02 \\
1359$-$0217 & 3657725094236611072 & He & 16773$\pm$55 & 8.14$\pm${0.02} & \hyperlink{24}{24} & 14$\pm^{4}_{3}$ & 1.4$\pm^{0.6}_{0.4}$ &   &   &   & 0.18 \\
J0644$-$0352 & 3105360521513256832 & He & 17000$\pm$327 & 7.98$\pm${0.02} & \hyperlink{7}{7} & 7$\pm^{2}_{2}$ & 6.3$\pm^{4.4}_{3.2}$ & 64.6 & 50.7$\pm^{7.3}_{7.3}$ & 0.16$\pm^{0.01}_{0.01}$ & 0.02 \\
GD\,61 & 203931163247581184 & He & 17280$\pm$400 & 8.20$\pm${0.30} & \hyperlink{29}{29} & 7$\pm^{4}_{3}$ & 2.1$\pm^{6.1}_{1.5}$ &   &   &   & 0.00 \\
WD\,1415+234 & 1253863445200890112 & He & 17300$\pm$500 & 8.17$\pm${0.05} & \hyperlink{25}{25} & 42$\pm^{10}_{9}$ & 1.2$\pm^{0.8}_{0.4}$ & -4.9 & 20.0$\pm^{9.2}_{9.2}$ & 0.28$\pm^{0.17}_{0.12}$ & 0.12 \\
SDSS\,J2248+2632 & 1883678961315891584 & He & 17370$\pm$500 & 8.02$\pm${0.05} & \hyperlink{25}{25} & 23$\pm^{8}_{7}$ & 4.3$\pm^{5.0}_{2.4}$ & -9.4 & 13.2$\pm^{6.9}_{6.9}$ & 0.08$\pm^{0.03}_{0.02}$ & 0.06 \\
WD\,J2047$-$1259 & 6888044253249541120 & He & 17970$\pm$155 & 8.04$\pm${0.05} & \hyperlink{30}{30} & 8$\pm^{4}_{3}$ & 3.5$\pm^{4.9}_{1.9}$ &   &   &   & 0.05 \\
Ton\,345 & 689352219629097856 & He & 19780$\pm$250 & 8.18$\pm${0.20} & \hyperlink{31}{31} & 43$\pm^{15}_{14}$ & 1.9$\pm^{5.4}_{1.2}$ & -3.7 & 22.6$\pm^{6.9}_{6.9}$ & 0.12$\pm^{0.08}_{0.06}$ & 0.06 \\
WD\,1536+520 & 1595298501827000960 & He & 20800$\pm$800 & 7.96$\pm${0.10} & \hyperlink{32}{32} & 21$\pm^{8}_{6}$ & 4.8$\pm^{5.2}_{3.0}$ &   &   &   & 0.14 \\
WD1622+587 & 1623866184737702912 & He & 21530$\pm$313 & 7.98$\pm${0.03} & \hyperlink{7}{7} & 28$\pm^{8}_{7}$ & 5.5$\pm^{4.8}_{2.9}$ & -51.8 & 124.8$\pm^{2.6}_{2.6}$ & 0.47$\pm^{0.12}_{0.12}$ & 0.13 \\
\enddata
\tablecomments{\hypertarget{a}{a}: 1: \citealt{Zuckerman2011}, 2: \citealt{LeBourdais2025}, 3: \citealt{Swan2019}, 4: \citealt{Xu2014}, 5: \citealt{Trierweiler2025}, 6: \citealt{Rogers2025}, 7: \citealt{Rogers2024}, 8: \citealt{Xu2019}, 9: \citealt{Melis2017}, 10: \citealt{Gansicke2012}, 11: \citealt{Williams2025}, 12:  \citealt{Melis2011}, 13: \citealt{Sahu2025}, 14: \citealt{Hollands2022}, 15: \citealt{Swan2023}, 16: \citealt{Izquierdo2026}, 17: \citealt{Zuckerman2007}, 18:  \citealt{Xu2013}, 19: \citealt{Klein2011}, 20: \citealt{Raddi2015}, 21: \citealt{Klein2021}, 22: \citealt{Dufour2012}, 23: \citealt{Xu2017}, 24: \citealt{Izquierdo2023}, 25: \citealt{Doyle2023}, 26: \citealt{Jura2012}, 27: \citealt{LeBourdais2024}, 28: \citealt{Izquierdo2021}, 29: \citealt{Farihi2013}, 30: \citealt{Hoskin2020}, 31: \citealt{Wilson2015}, 32: \citealt{Farihi2016} \\
\hypertarget{b}{b}: CMF (core-mass fraction) calculated as the metal/silicate ratio of the polluting material (Section \ref{section:CMF_calculation}). Values for H-WDs are steady-state adjusted. \\
\hypertarget{c}{c}: Age calculated as the sum of the cooling age of the WD and the main sequence lifetime (Section \ref{section:ages})\\
\hypertarget{d}{d}: $v_{\mathrm{R}}$ is the radial velocity calculated as the mean velocity shift of the photospheric metal lines reported by each reference, minus the gravitational redshift (Section \ref{section:methods_kinematics}). We uniformly apply an uncertainty of $\pm$15 km/s to account for typical variations in velocity offsets between lines, instruments, and the optical and UV measurements. \\
\hypertarget{e}{e}: $\mathrm{R_{vel}}$ is the distance in galactic velocity coordinates ($V$ and $\sqrt{U^2 + W^2}$) from the Sun (Section \ref{section:methods_kinematics}) \\
\hypertarget{f}{f}: $\mathrm{Z_{max}}$ is the maximum vertical excursion of each WD above the plane of the Milky Way, modeled from Gaia kinematics and the radial velocities reported in this table (Section \ref{section:methods_kinematics}). Values are only reported for WDs with reported velocity constraints. \\
\hypertarget{g}{g}: For WDs without $v_{\mathrm{R}}$, Min $\mathrm{Z_{max}}$ is obtained through iterative sampling of radial velocities (Section \ref{section:methods_kinematics}). }
\end{deluxetable*}

\subsection{MS Star Sample from APOGEE}

Throughout this work, we compare our results for the WD sample with abundance patterns in MS stars drawn from the APOGEE survey (Milky Way Mapper, DR19\citealt{SDSSCollaboration2025}).
We select stars with $4 < \log g \mathrm{~(cm/s^2)} < 6$ to restrict the sample to main-sequence stars. We additionally constrain our sample to targets with $2000<\mathrm{T_{eff}}/\rm{K}<7000$, $\mathrm{SNR>50}$, $\mathrm{Dist/Dist_{err}}>5$, $\mathrm{plx}>0$, $\mathrm{[Mg/H]}<2$, $\mathrm{[Fe/H]}<2$, and no ``BAD" flags, in accordance with the ASCAP guidelines \citep{Meszaros2025}. 
The resulting sample includes about 200,000 stars. 
For each star, we draw photometric and kinematic parameters from the Gaia catalog \citep{GaiaCollaboration2023}.  
The mass distribution for the MS sample is peaked at $0.9~\rm M_\odot$, relatively lower than the fitted progenitor masses for our WD sample, which range about $1-3~\rm M_\odot$ (see Section \ref{section:ages} for fitting methods). 

Uncertainties in abundances reported by APOGEE are based on the median scatter in repeated abundance measurements for single stars. 
These values capture systematic uncertainties within APOGEE but not necessarily calibration uncertainties across catalogs, such as those reported by the Hypatia Catalog \citep{Hinkel2014}. 
In order to draw comparisons between the MS and WD samples, we therefore adopt the minimum of the upper limit uncertainties calculated by \cite{Meszaros2025}. 
These estimates are reported relative to metallicity ([Z/M]) with an error of 0.035 subtracted in quadrature, and we therefore add this uncertainty back in to recover errors relative to [Z/H]. 

Finally, we classify the MS stars in our sample as either high- or low- alpha based on the criteria reported by \cite{Weinberg2019}, where high-alpha stars have ratios of $\mathrm{[Mg/Fe] > (0.12 - 0.13\times[Fe/H]})$ for $\mathrm{[Fe/H] < 0}$ and $\mathrm{[Mg/Fe ]> 0.12}$ for $\mathrm{[Fe/H] > 0}$. 

\section{Methods}\label{section:methods}

\subsection{Calculation of iron mass fractions from WD and stellar abundances}\label{section:CMF_calculation}

We calculate the mass ratio of iron metal to total (iron and silicate) material, a proxy for an expected core-mass fraction of the assembled body, as 
\begin{equation}\label{eq:fFe}
     \mathrm{CMF} = m_{\rm Fe} / \left(m_{\rm Fe} + m_{\rm Mg_2SiO_4} + m_{\rm MgSiO_3} + m_{\rm SiO_2}\right),
\end{equation}
\noindent where $m_i$ is the abundance by number $\mathrm{n_i}$ of the species $i$ multiplied by the formula weight. 
The number abundances of the silicate species are calculated by solving 
\begin{equation}
\begin{split}
    &\mathrm{n_{Mg} = 2 \times n_{Mg_2SiO_4} + n_{MgSiO_3}} \\
    &\mathrm{n_{Si} = n_{Mg_2SiO_4} + n_{MgSiO_3} + n_{SiO_2}},
\end{split}
\end{equation}

\noindent so that $\mathrm{n_{MgSiO_3} =2 n_{Si}-n_{Mg}}$, $\mathrm{n_{Mg_2SiO_4}=n_{Mg}-n_{Si}}$, and $\mathrm{n_{SiO_2}}$ is any remaining Si. 
$\mathrm{n_{Fe}}$ is calculated directly from the measured Fe abundance.
We assume all Fe is present as metal, as opposed to FeO, and correspondingly assume no FeO is in the silicate portion of the material. Applying an Earth-like concentration of FeO can reduce CMF by a few percent (within $1\sigma$ uncertainties of the values we report), while more extreme oxidation conditions would render the iron mass fraction to an upper limit of metal mass.
We propagate uncertainties in CMF from the measured abundance uncertainties using a bootstrap approach. For each target, we take 100,000 random draws in the log abundance of each measured element relative to H or He, propagate these values through Equation \ref{eq:fFe}, and report the resulting 16, 50, and 84th percentile quantities as the median and 1$\sigma$ uncertainty intervals. 

For the MS stars, we also tested calculating CMF while applying a devolatilization factor so that the element ratios are $\mathrm{Z/Mg|_{devol} = Z/Mg \times T_{cond, Z}/T_{cond, Mg}}$, where $\mathrm{T_{cond}}$ is the condensation temperature of each element. 
However, because the CMF depends on Mg, Si, and Fe, which have very similar condensation temperatures, the resulting variation in CMFs is within $1\sigma$ uncertainties. 

The mass of O in the silicate is constrained solely from the Mg and Si abundances, for both WDs and MS stars, because O excesses in both cases are common and thus do not put strong constraints on silicate material. O excess in WDs can result from water accretion \citep[e.g.,][]{Trierweiler2025}, while nucleosynthetic O production in stars, combined with O's relative volatility, make it unlikely for stellar O/silicate ratios to correspond directly to those of the resulting planetary material.
For the WDs, all but five samples have at least enough O to transform all observed Mg and Si into oxides. 
The samples lacking enough O are GD56, PG0843+516, PG1015+161, and WD1622+587. 
GD56, PG0843+516, and PG1015+161 are H-dominated, and PG0843+516, and PG1015+161 do have a sufficient amount of O to account for Mg and Si oxides when steady-state is applied. 
On an individual basis, invoking different accretion phases or extreme disk timescales can help to reduce the O gaps, however a more thorough study of WD systems lacking O relative to Mg and Si would be useful to identify whether a physical mechanism or systematic in modeling could be responsible for the discrepancies. 

In calculating CMFs, we additionally apply a steady-state adjustment to the H-dominated WDs, based on the assumption that these systems must be in or near steady-state to be observed given their short settling timescales relative to accretion timescales. For each element in Equation \ref{eq:fFe}, we apply the transformation 
\begin{equation}
    \mathrm{n_{Z, SS} = n_{Z, obs} \times \frac{\tau_{\mathrm{d}} - \tau_{\mathrm{Z}}}{\tau_{\mathrm{Z}}}} \left( e^{-t_{\mathrm{SS}}/\tau_{\mathrm{d}}} - e^{-t_{\mathrm{SS}}/\tau_{\mathrm{Z}}} \right)^{-1}
\end{equation}\label{Eq:nSS}

\noindent to Fe, Si, and Mg abundances before calculating the CMF. Here $\mathrm{n}$ is the abundance of the element in WD atmosphere by number, $\tau_{\mathrm{d}}$ is the characteristic lifetime of the accretion disk, $\tau_{\mathrm{Z}}$ is the settling timescale for the element, and $t_{\mathrm{SS}}$ is the time after the onset of accretion that the system reaches steady state. 
We assume $\mathrm{\tau_{\mathrm{d}} = 10^{5.6}~yr}$ to represent a typical WD accretion disk, as constrained by \cite{Cunningham2021}. Settling timescales are interpolated for each WD using the grids from the Montreal White Dwarf Database \citep{Dufour2017}. Under the exponential accretion and settling model \citep{Jura2009}, for a given element steady state accretion is reached at the time $\mathrm{\frac{\tau_{\mathrm{d}} \tau_{\mathrm{Z}}}{\tau_{\mathrm{d}}-\tau_{\mathrm{Z}}} \ln(\tau_{\mathrm{d}}/\tau_{\mathrm{Z}})}$, and we define $t_{\mathrm{SS}}$ as the median steady state time of the elements considered (Mg, Si, and Fe). 

\subsection{Extrapolation of WD and stellar properties}
To test for composition variations over time, we require a tracer of age for the WDs and MS stars. We therefore obtain stellar age and kinematic estimates for each object in our sample. 

\subsubsection{Stellar age}\label{section:ages}
For the WDs, we constrain total stellar ages from the present-day mass and cooling age determined for each object. 
Using the effective temperature and $\log g$ parameters reported in the literature (Table \ref{table:WDsample}), we calculate the total age using \textsc{wdwarfdate} \citep{Kiman2022}, which includes cooling models from \citealt{Bedard2020} and isochrones from \citealt{Choi2016} and \citealt{Dotter2016}.
In these fits, we use the initial-final mass relation (IFMR) from \citealt{Cummings2018}, and assume solar metallicities for the WD progenitors. 
The resulting ages include uncertainties from effective temperature and $\log g$, propagated from the uncertainties listed in Table \ref{table:WDsample}. 
For samples without reported uncertainties in WD parameters, we instead input values of 500~K and 0.04~dex for $\mathrm{T_{eff}}$ and $\log g$, respectively, representing typical parameter uncertainties \citep{Doyle2023}.  
\textsc{wdwarfdate} does not attempt to estimate ages if the fitted WD mass is outside of the initial-final mass relation (with a lower limit around $0.5~\mathrm{M_\odot}$) and/or outside of the typical range of masses for single star evolution ($0.45-1.1~\mathrm{M_\odot}$).

Uncertainty in total age reported by \textsc{wdwarfdate} for the WDs is generally very large, reaching up to about 4~Gyr for the oldest stars. 
Amongst the IFMRs included in \textsc{wdwarfdate}, the \citealt{Cummings2018} MIST and PARSEC, and \citealt{Marigo2020} functions return total ages well within uncertainty limits of each other. 
The \citealt{Salaris2009} and \citealt{Williams2009} likewise return total ages within uncertainty limits, though typically find much lower ages than those from \citealt{Cummings2018} and \citealt{Marigo2020}. 
Nonetheless, the relative ages of the WDs stay relatively similar between different IFMRs, with the oldest and youngest WDs remaining at the extremes of the age sample. 
For the remainder of this work, we use the \citealt{Cummings2018} IFMR results, as recommended by \citealt{Kiman2022}. 

We also consider the impact of variations in reported WD parameters, in particular for systems where the initial abundance fits asserted a particular temperature or $\log g$ value. 
For this purpose, we collect the most recently reported WD parameters from the Montreal White Dwarf Database \citep[MWDD, ][]{Dufour2017}, and carry out the same process of calculating ages with \textsc{wdwarfdate}.
The corresponding values and references are listed in Table \ref{table:MWDD_recent_WDsample}. 
The resulting total ages are typically within $1\sigma$ of those calculated from the parameters reported in Table \ref{table:WDsample}. 
Additionally, some WDs with invalid ages using the parameters reported in Table \ref{table:WDsample} have valid \textsc{wdwarfdate} fits using the updated MWDD parameters. 
We carry out separate analyses for ages derived from reported parameters versus the recently reported MWDD parameters (``MWDD-Recent"). 
In general, the age uncertainties calculated for the WDs tend to be $30\%$ or greater, exceeding the age uncertainties of the MS stars which are typically around $10\%$.
Note that the MWDD-Recent parameters often come from larger surveys, and the abundances of the pollutants are not remodeled with the updated parameters.
While WD parameters do impact abundance fits, the relative abundances of Si, Mg, and Fe used in this work to calculate CMF values should be robust to changes in $\mathrm{T_{eff}}$ of up to a few thousand Kelvin and in $\log g$ of about 0.3~dex \citep{Klein2011, Rogers2024a}, which is typically within range of the parameter changes from reported to MWDD-recent values.  

Using the \citealt{Cummings2018} MIST IFMR, the majority of WD masses range from about $0.55-0.77~\mathrm{M_\odot}$. 
Three WDs (G29$-$38, 0944$-$0039, and 1516$-$0040) have masses calculated by \textsc{wdwarfdate} between $0.8-1.0~\mathrm{M_\odot}$, when using the WD parameters reported in Table \ref{table:WDsample}. 
These WDs have a $10-20\%$ probability of being merger remnants \citep{Cheng2020}, and thus may have ages that are underestimated compared to their kinematics. 
However, in all three cases the most recent parameters from the MWDD return masses between $0.5-0.7~\mathrm{M_\odot}$. 

For the MS stars, we first draw [Fe/H] as well as $\mathrm{T_{eff}}$ and $\log g$ from APOGEE, along with associated uncertainties. For each target, we also select parallax and G, $G_{\rm BP}$, and $G_{\rm RP}$ photometry from the Gaia catalog. 
We then use the stellar classification package \textsc{isoclassify} \citep{Huber2017, Berger2020} to fit the age of each star, using the `allsky' dustmap \citep{Bovy2016} and MESA grid \citep{Choi2016}. 
For successful fitting of ages, we apply uniform uncertainties of 0.01 to reported Gaia photometry. 
About 8000 of our selected 130000 return ages of zero or lower age limits $>14$~Gyr. 
These are typically flagged as M-dwarfs by \textsc{isoclassify}, and we remove these stars from our sample. 

\subsubsection{Galactic kinematics}\label{section:methods_kinematics}
As a second tracer of stellar evolution, we independently draw kinematic data for each WD and MS star from Gaia, and calculate the U, V, W velocity components and vertical amplitude ($\mathrm{Z_{max}}$) of each star's oscillations through the galactic disk. 

For this calculation, we transform the RA, Dec, proper motions, and radial velocity to galactic coordinates with \textsc{astropy} \citep{AstropyCollaboration2022}. 
Reported U, V, W velocities are the calculated x, y and z velocities from \textsc{astropy}, respectively, and are not adjusted to the local standard of rest. 
To carry out tests with kinematic velocities, we defined the quantity $R_{\rm vel}$ as the distance, in the $V$ and $\sqrt{U^2 + W^2}$ space, of a given star from the Solar values, 
\begin{equation}\label{eq:Rvel}
    R_{\rm vel}^2 = (\sqrt{U^2 + W^2} - \sqrt{U_\odot^2 + W_\odot^2})^2 + (W-W_\odot)^2,
\end{equation}
\noindent where $U_{\odot} = 240$~km/s \citep{Reid2014}, $V_{\odot} = 11.1$, and $W_{\odot} = 7.25$ \citep{Schonrich2010}. 

\begin{figure}
    \centering
    \includegraphics[width=1\linewidth]{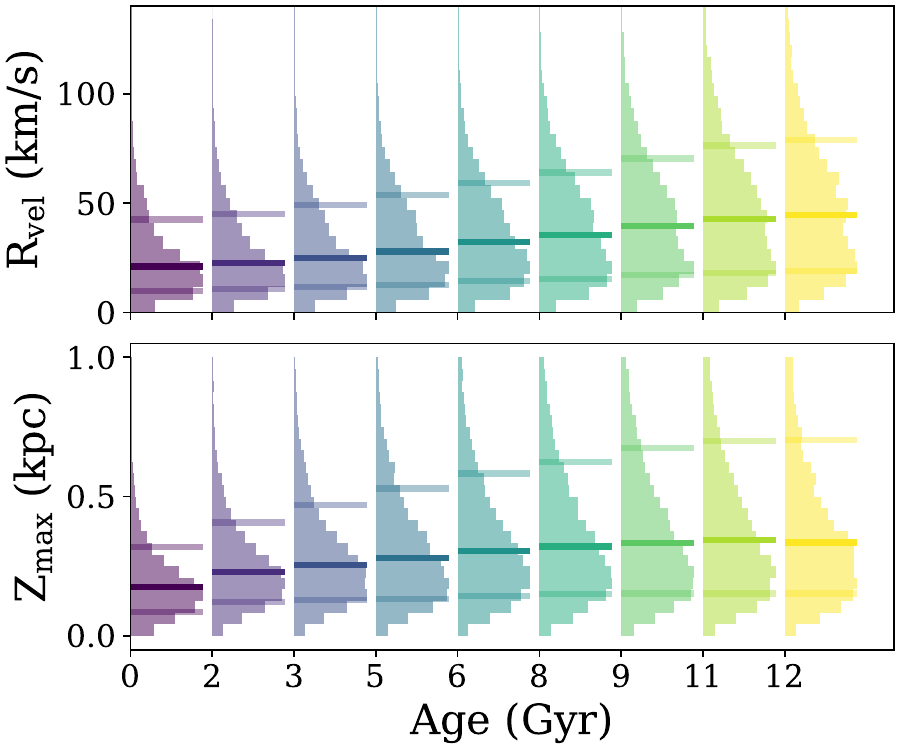}
    \caption{Distribution of $\mathrm{R_{vel}}$ (Equation \ref{eq:Rvel}) and $\mathrm{Z_{max}}$ (the maximum vertical amplitude) of MS stars of increasing ages. The horizontal lines show the median and $1\sigma$ values for each age bin. Younger stars tend to have smaller $\mathrm{Z_{max}}$ values while older stars trend toward larger amplitudes and wider distributions of $\mathrm{Z_{max}}$. }
    \label{fig:zmax_age_apogee}
\end{figure}

Using the calculated velocities, we then integrate the resulting orbits with the \textsc{Gala} package \citep{Price-Whelan2017} to solve for $\mathrm{Z_{max}}$.
We validate the use of kinematic data as a rough proxy for age in Figure \ref{fig:zmax_age_apogee}. 
For the MS star sample, the distribution of maximum vertical amplitudes increases and widens with stellar age, consistent with stars being dynamically excited over their lifetimes \citep[e.g.,][]{McCluskey2024}.
Comparison between age and $\mathrm{R_{vel}}$ gives very similar results. 

Radial velocity measurements for the WDs are difficult to obtain due to pressure broadening of spectral lines \citep[e.g.,][]{Saumon2022}, and are typically not reported by \textit{Gaia}.
Where available, we instead estimate radial velocity by subtracting the calculated gravitational redshift for each WD (interpolated from the MWDD) from its mean heliocentric velocity reported in the literature (Table \ref{table:WDsample}). 
The heliocentric velocities are the measured shifts in the observed photospheric metal lines. 
We apply an uncertainty of 15~km/s to the estimated radial velocities to encompass typical variation in measured velocities between different elements and between the UV and optical bands. 
We then propagate these radial velocities through the integrations to obtain vertical amplitude estimates. 
For WDs with no reported velocity shifts, we additionally calculate a minimum estimate for the maximum vertical amplitude by iteratively sampling radial velocities between $-250$ and $250$ km/s, integrating the resulting orbit, and selecting the minimum vertical extent for each WD. 

\subsection{Characterizing trends in iron mass fraction vs stellar age}\label{methods:metal_age_test}

GCE suggests that planets formed around younger stars should have systematically higher CMFs than those formed in older systems. 
We test for this trend in two ways. 

First, we carry out a linear fit to the CMF and age/kinematic data using bootstrapping to account for uncertainties. 
Using the distributions of the fitted slopes, we report a probability that CMFs follow a negative slope relative to the total age of the system. 
We select linear fits as opposed to more complex models as our goal is to test to first-order positive or negative correlations between CMF and system age. 

Second, we split the WD and MS star samples into young and old populations for various threshold ages and vertical amplitudes and compare the median CMFs of the two populations. 
In this approach, we use bootstrapping to propagate errors in core-mass fraction and age to calculate the 16, 50, and 84th percentile values of the median CMFs for the young and old populations. 
We then use the distribution of median CMF values for the two populations to test for a significant difference between planetary material in young and old systems. 

\subsection{Population synthesis of expected iron mass fraction trends}\label{section:methods_popsynth}
To estimate the significance of CMF disparities between old and young WD systems that can be achieved as a function of sample size, we carry out a population synthesis model, propagating the synthetic data through the analysis of Section \ref{methods:metal_age_test}. 

For each simulated data point, we draw a stellar age at random from 0.5 to 14~Gyr. 
We then generate a corresponding CMF value by randomly sampling the distribution of CMF values calculated for the MS star sample, using a slice of MS stars spanning ages within 0.1~Gyr of the selected age. 
In addition to this process, we impose uncertainty terms informed by the observed polluted WDs, $\mathrm{\sigma_{age}/age}$, $\mathrm{\sigma_{acc}}$ and $\mathrm{\sigma_{abund}}$. 

The first component of uncertainty, $\mathrm{\sigma_{age}/age}$, accounts for the uncertainty in determining stellar ages. 
Uncertainties tend to scale with stellar age, largely as a result of the initial-final mass relation for WDs, so we implement this term as a fractional uncertainty. 

The second component of uncertainty, $\mathrm{\sigma_{acc}}$, considers the degree to which the observed accretion phase of the WD could skew the measured CMF. 
For each WD, neglecting measured abundance uncertainties, we calculate the variation in CMFs at a range of times around steady-state (Section \ref{section:CMF_calculation}). To ensure a non-negligible amount of pollution is present in the atmosphere, we choose a time range spanning from $\mathrm{0.2\times min(\tau_{\mathrm{Z}}, \tau_{\mathrm{d}}})$ to $\mathrm{5\times max(\tau_{\mathrm{Z}}, \tau_{\mathrm{d}}})$, where $\tau_{\mathrm{Z}}$ and $\tau_{\mathrm{d}}=10^{5.6}$ yr are the average per-element settling timescale and assumed disk lifetime, respectively. 
We then calculate $\mathrm{\sigma_{acc}}$ for the H and He WDs as the median uncertainty in the CMF for each sample. 
While this is not an exhaustive account of accretion physics that could affect observed CMF, $\mathrm{\sigma_{acc}}$ represents a minimum estimate of the variation in CMF that could be inherent to sampling the polluted WDs at different times relative to the onsets of their accretion events. 

The final imposed component of uncertainty, $\mathrm{\sigma_{abund}}$, is the variation in CMF directly resulting from measurement uncertainty in the log abundance ratios of individual elements in the WD atmosphere. 
We first propagate uncertainties in the log abundances for each WD through Equation \ref{eq:fFe} (ignoring the impact of accretion physics). 
We then calculate $\mathrm{\sigma_{abund}}$ as the 84th percentile of uncertainty in CMF for the H and He WD samples. 

Overall, we find values of $\mathrm{\sigma_{age}/age}=0.3$, $\mathrm{\sigma_{acc}}=$0.0, 0.16 and $\mathrm{\sigma_{abund}}=$0.13, 0.08, for the H and He samples, respectively. 

In each instance of the population synthesis model, we select a sample size, WD atmosphere type (H or He), and limiting age to separate old and young WDs.
For each synthetic WD, we choose an age at random and draw a ``true" CMF from the CMF distribution of MS stars. 
We then generate an accretion-phase-dependent CMF value by sampling a random normal distribution where the mean is defined by the linear CMF-age relation and the standard deviation is $\mathrm{\sigma_{acc}}$. 
We draw an ``observed" CMF from a random normal distribution where the mean is the accretion-dependent value and the standard deviation is $\mathrm{\sigma_{abund}}$. 
Finally, we draw a ``measured" age for the synthetic WD by sampling a normal distribution with a mean of the selected stellar age and a standard deviation of that age multiplied by $\mathrm{\sigma_{age}}$, and use this measured age to categorize the generated WD as young or old. 
As for the observed WDs, we iterate this process to calculate the median and uncertainties of the  CMFs for the young and old populations.

\section{Results}\label{section:results}

\subsection{Baseline trends in iron mass fractions from MS stars}

\begin{figure}
\centering
\includegraphics[width=\linewidth]{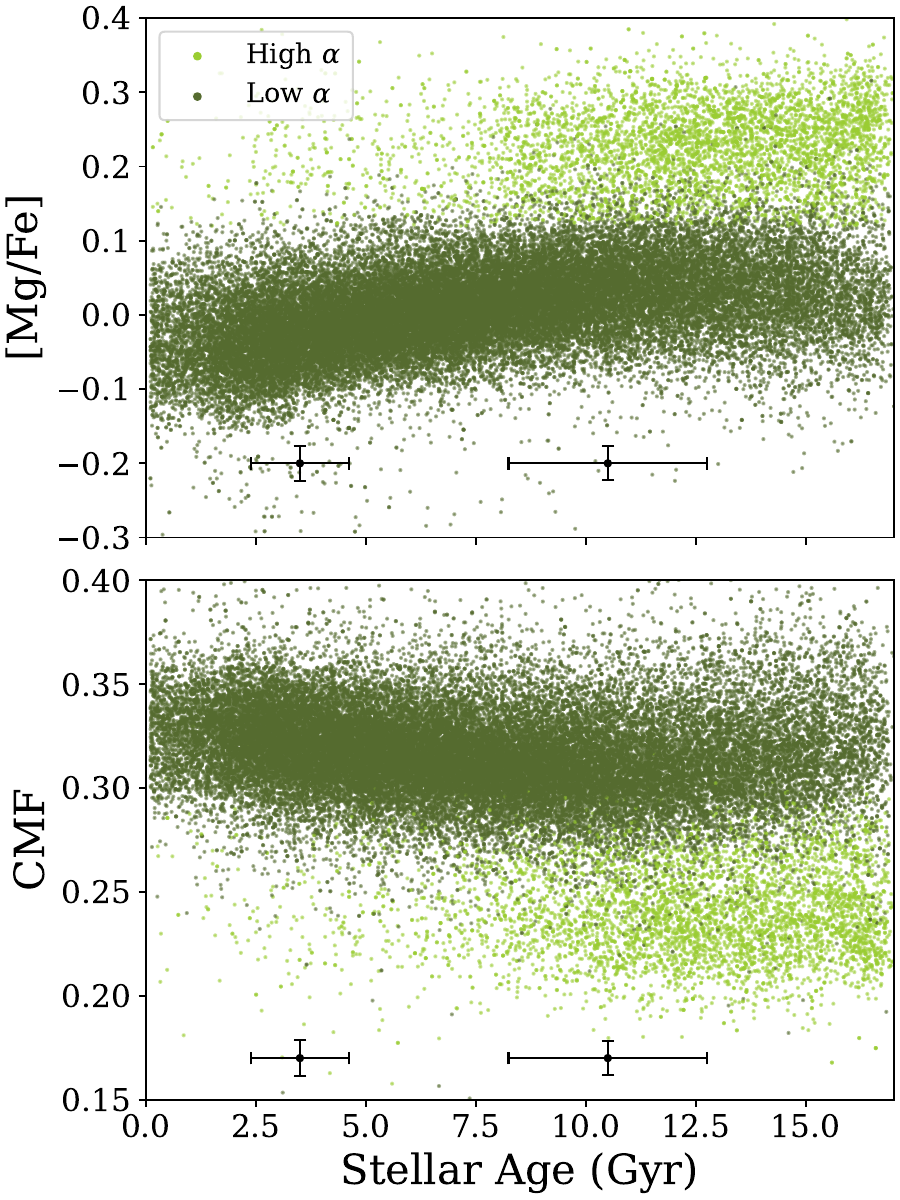}
\caption{Abundance ratios and calculated iron mass fractions for MS stars drawn from APOGEE, relative to fitted stellar ages (Section \ref{section:ages}). Points in light green are part of the  high alpha-element population, while those in dark green belong to the low-alpha-element population. The plotted error bars show median per-point uncertainties for old ($\gtrsim 10~\mathrm{Gyr}$) and young ($\lesssim 10~\mathrm{Gyr}$) stars. }
\label{fig:comp_age_apogee}
\end{figure}

Figure \ref{fig:comp_age_apogee} shows the [Mg/Fe] ratios of the MS stars, representing relative lithophile/siderophile element ratios, as well as the calculated CMFs (Section \ref{section:CMF_calculation}) as a function of the fitted stellar age.  
The trend in [Mg/Fe] shows the motivation for the expectation that older stars should form bodies with relatively smaller metal cores than those formed around younger stars. 
Translating the stellar abundances to a CMF, we find that CMF may vary between 0.2 for the oldest, high-alpha element stars, to nearly 0.4 for the younger, low-alpha element stars. 
These trends do not show direct proportions between changes in age and Mg/Fe or CMF, but demonstrate that because older stars can be part of the high-alpha element group, they have some likelihood of hosting low CMF material. 

\begin{figure}
    \centering
    \includegraphics[width=\linewidth]{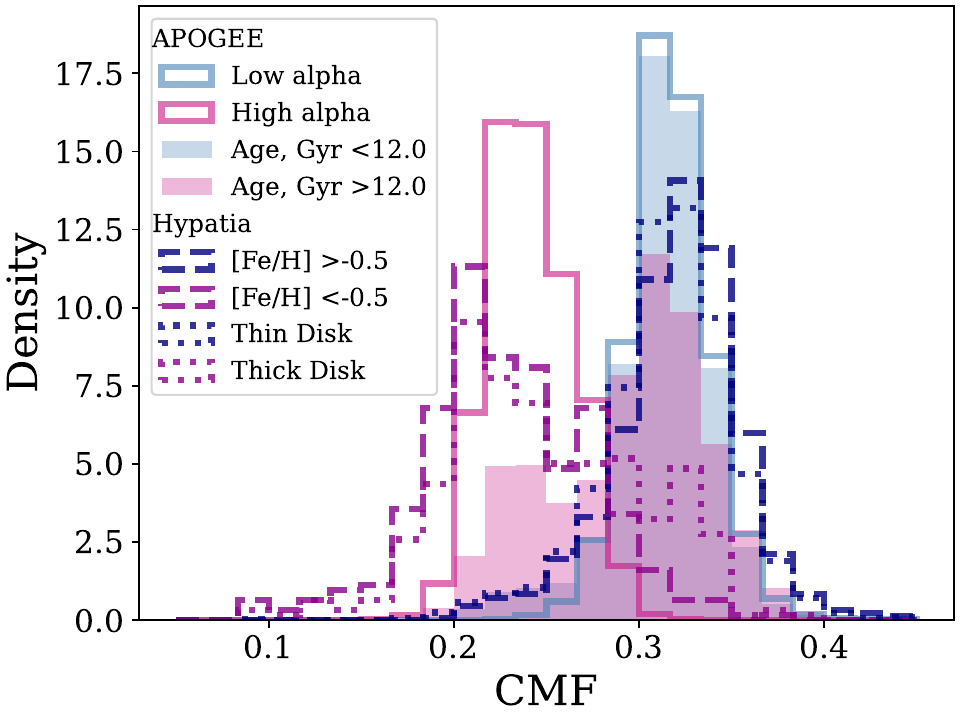}
    \caption{Distribution of CMF values for MS stars drawn from APOGEE and Hypatia, grouped by stellar properties. The high and low metallicity and thin/thick disk Hypatia curves (see \citealt{Michel2020, Trierweiler2023}) are well matched by the high and low alpha-element samples from APOGEE. Grouping stars by stellar age (shaded regions) still results in variations in CMF, though there is much more overlap between the young and old groups. }
    \label{fig:apogee_hypatia_cmf_compare}
\end{figure}

In Figure \ref{fig:apogee_hypatia_cmf_compare}, we compare the distribution of CMF values for the MS stars drawn from APOGEE to those reported in the Hypatia Catalog \citep{Hinkel2014}, representing previous works \citep{Michel2020, Trierweiler2023}. 
The variations between CMF for different populations predicted in previous works is largely dependent on the correspondingly low and high CMF values for high and low-alpha MS stars. 
Applying a metallicity threshold of [Fe/H]$=-0.5$ typically well separates the high and low-alpha stars. 
However, such a clean split in age is more difficult. We arbitrarily show CMF distributions for old and young stars defined with a threshold age of 12 Gyr, an extreme threshold chosen to illustrate the age/metallicity overlap. The result produces a bimodal distribution for the old (pink) population, which contains both high and low-alpha stars.  

These patterns in MS stars place bounds on how strongly CMF could vary with age tracers for the WD sample. 
Assuming WDs roughly follow metallicity and alpha-element trends reflected in MS stars, we expect older WDs to have a larger proportion of low-CMF material. 
This should manifest as a small negative slope in CMF with age, and a relatively lower CMF amongst older WDs, though fully separating low and high CMF populations is unlikely using only stellar age as a tracer. 

\subsection{Iron mass fractions of WD pollution}\label{section:results_cmf_wd}

\begin{figure}
    \centering
    \includegraphics[width=1\linewidth]{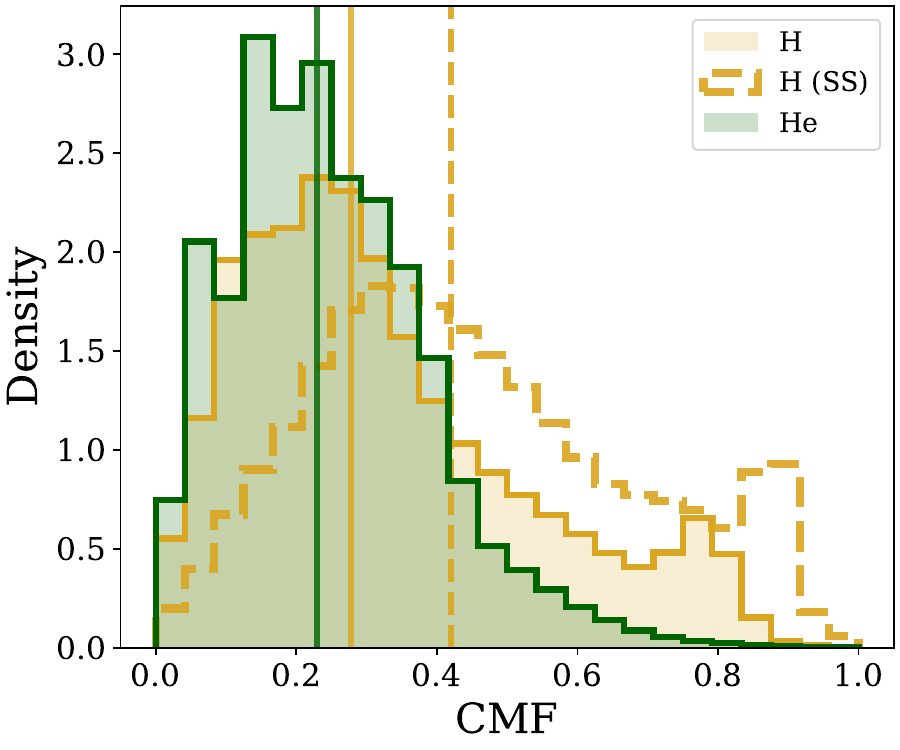}
    \caption{Calculated iron mass fractions for the polluted WD sample, grouped by the dominant element in the WD atmosphere. For comparison, the Earth's CMF is about 0.3. The vertical lines show the median values for each subset of WDs. For H-WDs, we also show iron mass fractions assuming steady-state accretion (SS). The steady-state assumption tends to result in larger iron mass fractions, due to differential element settling. }
    \label{fig:wd_cmf_dist}
\end{figure}

Figure \ref{fig:wd_cmf_dist} shows the calculated CMFs for the WD sample, split based on the dominant element in the WD atmosphere (H or He). 
For the H-WDs, we show the CMFs calculated from the observed elements as measured, as well as under the assumption of steady-state accretion (Section \ref{section:CMF_calculation}). 
The WD CMFs peak between 0.2 and 0.4, roughly around the Earth's CMF of $\approx0.36$, using the bulk Earth composition of \citealt{McDonough2026}. This is broadly consistent with previous studies of core material in polluted WD systems, such as \citealt{Bonsor2020} and \citealt{Bonsor2023}.

Because steady-state occurs after several settling timescales for H WDs, the system has already undergone differential element settling. 
This tends to lead to a larger inferred amount of Fe relative to Si and Mg in the steady-state scenario, as Fe typically settles more quickly than other elements, and therefore results in larger CMFs. 
However, it is unlikely that material accreting onto H-WDs is inherently more Fe-rich than that accreted by He-WDs, and we discuss the potential explanations for the difference between the steady-state and observed distributions for H-WDs in Section \ref{section:HvHe_ratios}. 
For the remainder of the work, we report separate results for the H-WDs for both the observed and steady-state scenarios. 
The qualitative results for the two assumptions tend to be similar, with shifts in exact values corresponding to the relatively higher CMFs calculated under steady state. 

\subsection{Variation in iron mass fractions with stellar age}\label{section:results_cmf_age_fits}

\begin{figure}
    \centering
    \includegraphics[width=1\linewidth]{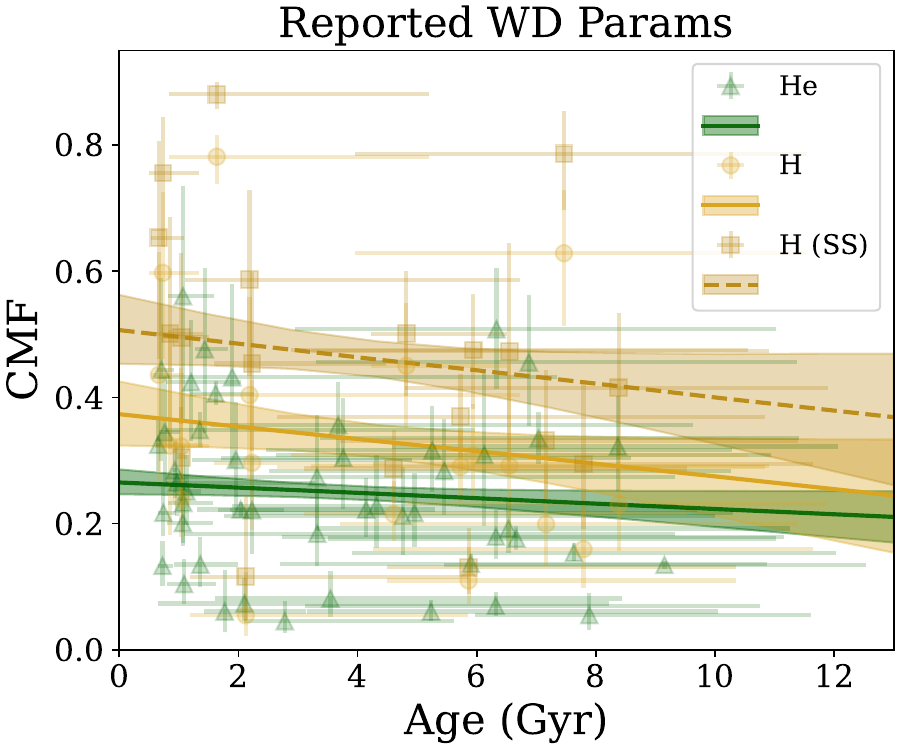}
    \includegraphics[width=1\linewidth]{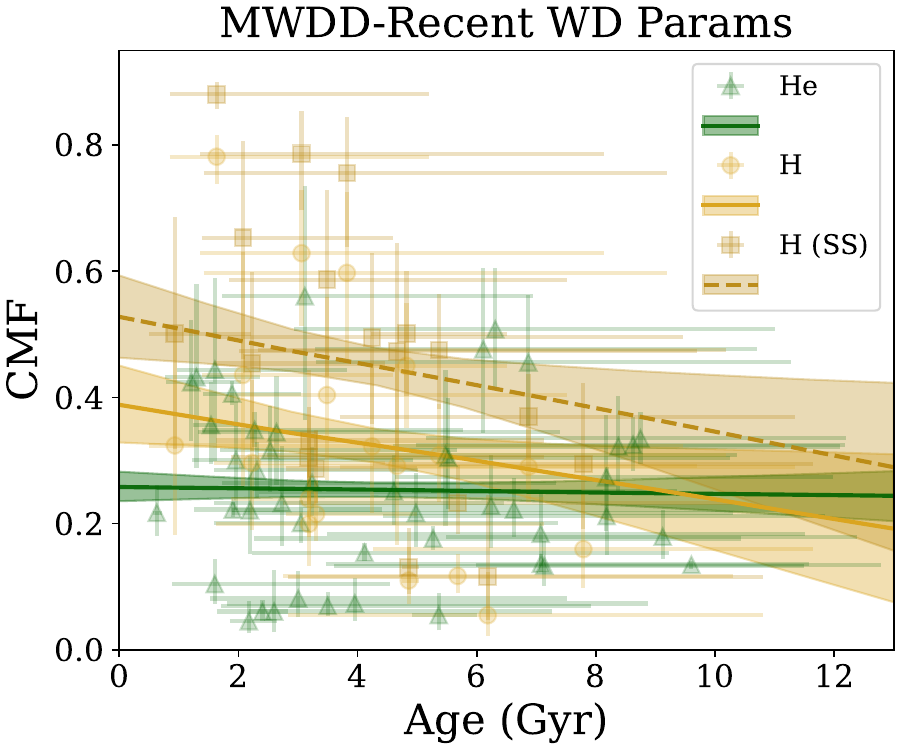}
    \caption{Calculated iron mass fraction versus total stellar age for the polluted WD sample, separated by atmospheric type (H vs He) and with a steady-state (SS) adjustment applied to the H-WDs.
    The top panel uses ages calculated from the WD temperature and $\log g$ values reported in the abundance fits for each WD, and the bottom panel uses the most recently calculated WD parameters, as reported by the MWDD. 
    The curves show the linear fits obtained through bootstrap fitting, with 1$\sigma$ uncertainties.
    }
    \label{fig:cmf_age_fit}
\end{figure}

Figure \ref{fig:cmf_age_fit} shows the WD CMFs as functions of the calculated total stellar age and vertical amplitude, and the resulting linear fits for each case (Section \ref{methods:metal_age_test}). 
Both the H and He samples exhibit negative slopes between CMF and age, though the trend is much shallower for the He-WDs. 
Considering the fraction of bootstrap samples with negative slopes as the probability of a negative trend in CMF with age, the He and H  (both as-observed and adjusted for steady state) samples have probabilities of negative slope of 85\% and 83\%, respectively. 

\begin{figure}
    \includegraphics[width=1\linewidth]{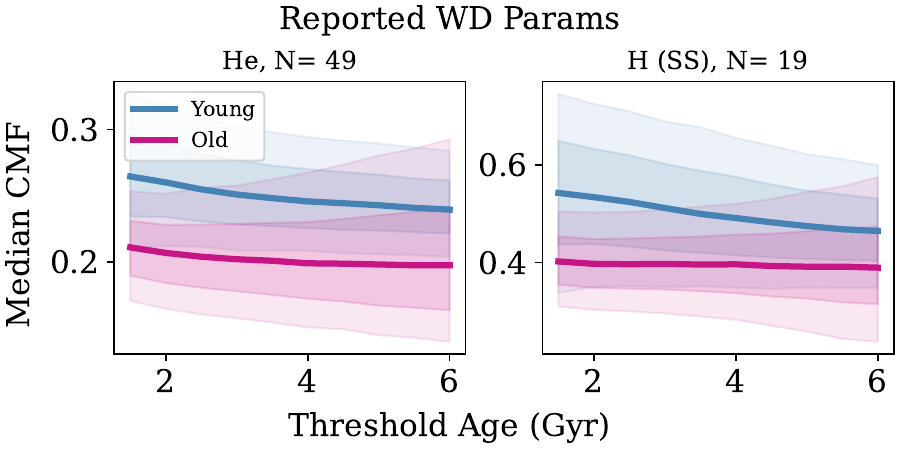}
    \includegraphics[width=1\linewidth]{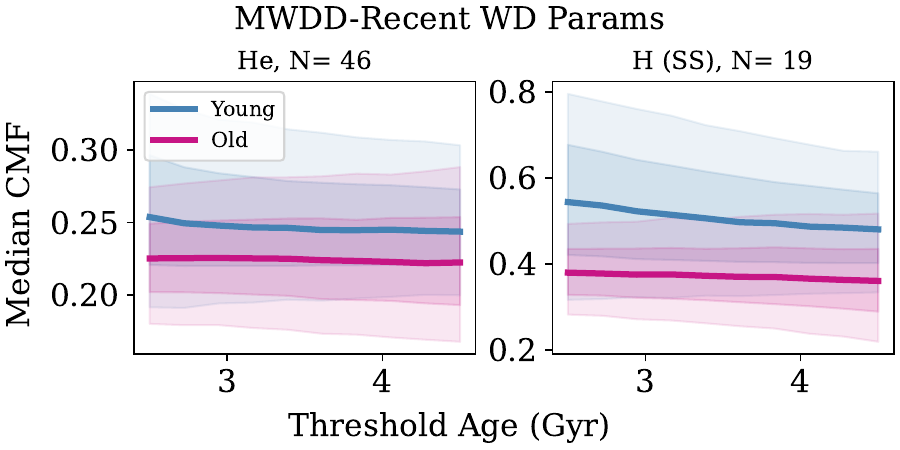} 
    \caption{The median iron mass fractions (CMF) for old (pink) and young (blue) populations of WDs, where old and young are defined relative to a range of threshold values. 
    Young populations are defined as being below the reported age threshold. 
    Each panel corresponds with a different WD type and lists the associated sample size. 
    Abundances for H-dominated WDs are adjusted to steady-state (SS).
    The shaded regions show the $1\sigma$ and $2\sigma$ values derived from bootstrapping. }
    \label{fig:linear_med_cmf}
\end{figure}

Figure \ref{fig:linear_med_cmf} shows the distribution of median CMFs for young and old polluted WD samples for different threshold ages, approximately spanning from the first to third quartiles of the WD age distribution. 
The young and old distributions include uncertainties in WD ages. 
For the H-WD sample, median CMFs for the young population remain higher than those of the old populations at about a 1$\sigma$ level for both the as-observed and steady state cases. 
The He-WDs reach a $1\sigma$ separation in CMF for young and old populations when assuming a threshold age below about 2 Gyr. While the median CMF of the young population remains higher than that of the old population at larger threshold ages, the difference is less than $1\sigma$. 

For some WDs in our sample, the effective temperature and $\log g$ used in the reported abundance fits differ from more recently reported parameters in the literature.  
This can be due to new survey coverage, updated observations, or variations in modeling methods (for example, with or without atmospheric metals).
We therefore also carry out the same analyses for total age using more recent fits for $\mathrm{T_{eff}}$ and $\log g$ (lower panels of Figure \ref{fig:linear_med_cmf}). 
For this, we use the stellar parameters most recently reported in the MWDD to carry out the total age calculations. 
For the H-WDs, we find relatively little change between the results using reported versus updated WD parameters. 
In both cases, CMF trends downwards with age with probabilities of negative slope of 84-87\%. 
While the He-WDs have relatively little change in CMF with age, the probability of negative slope is lower using the MWDD parameters ($60\%$) compared to that of the reported parameters ($84\%$). 

In Figure \ref{fig:linear_med_cmf}, we propagate uncertainties in CMFs from both abundance uncertainties and age. 
Neglecting the uncertainties in age increases the separations for the He sample, so that it reaches a maximum of $2\sigma$ for low age thresholds, as well as the separations for the H sample, which reach 2$\sigma$ and 3$\sigma$ for the low and high age thresholds, respectively. 

\subsection{Variation in iron mass fractions with galactic kinematics}\label{section:results_kinematics_fits}

\begin{figure}
    \centering
    \includegraphics[width=1\linewidth]{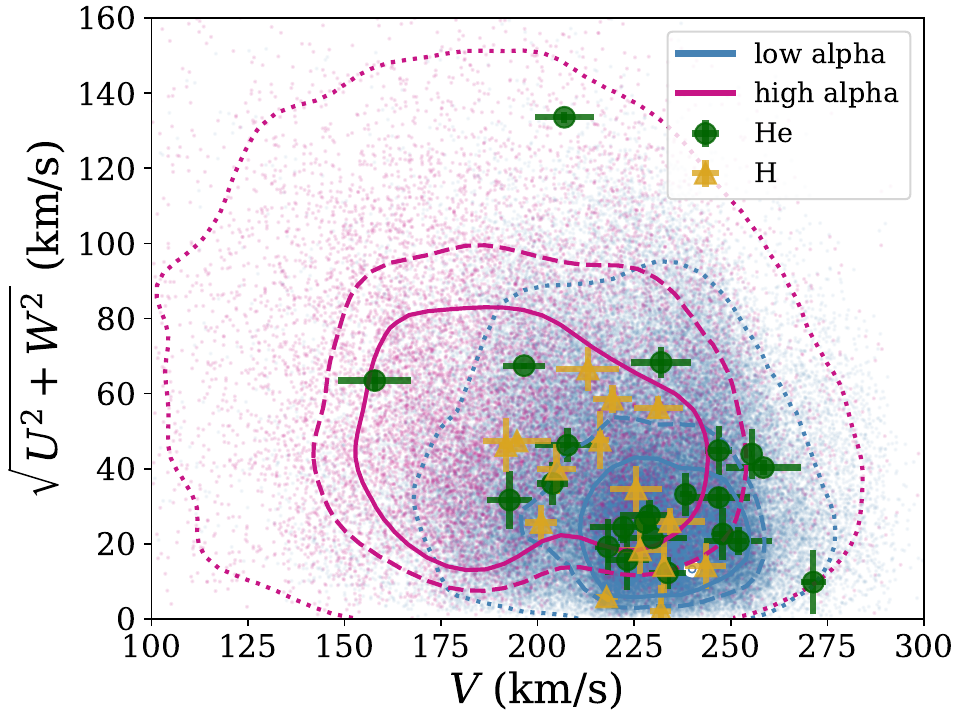}
    \caption{The polluted WD sample (large points) in kinematic space, compared to the MS sample (small points). 
    The Sun is indicated by the white star. 
    The WDs are divided by atmospheric type, and the MS stars are colored according to their classification as high or low $\alpha$/Fe stars. 
    The solid, dashed, and dotted curves show the 50, 68, and 95$\%$ levels for the MS star samples, respectively. }
    \label{fig:uvw_space}
\end{figure}

Figure \ref{fig:uvw_space} shows the calculated U, V, and W galactocentric velocities for the polluted WD sample, compared to the MS stars with high and low $\alpha$/Fe. 
While there is considerable overlap between the high and low $\alpha$/Fe stars in kinematic space, the high $\alpha$/Fe (relatively older) stars tend to be situated further from the Sun's velocity coordinates. 
The WDs are more closely clustered near the Sun in velocity space, but the sample does extend into the MS high $\alpha$/Fe region. 
Two He-WDs, G241$-$6 (far left side of figure) and WD~1622+587 (top of figure), have velocities that are particularly far from the low-$\alpha$/Fe distribution. 
Kinematically, these WDs are therefore consistent with being part of a thick disk or halo population, though this is not necessarily a guarantee that they formed with a high $\alpha$/Fe ratio.

\begin{figure}
    \centering
    \includegraphics[width=1\linewidth]{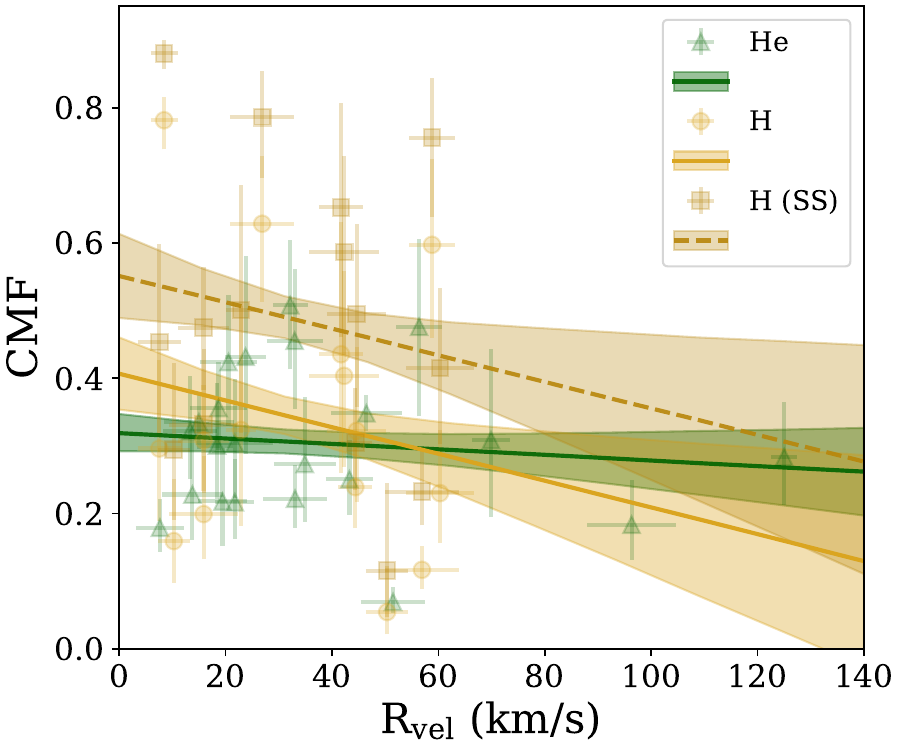}
    \includegraphics[width=1\linewidth]{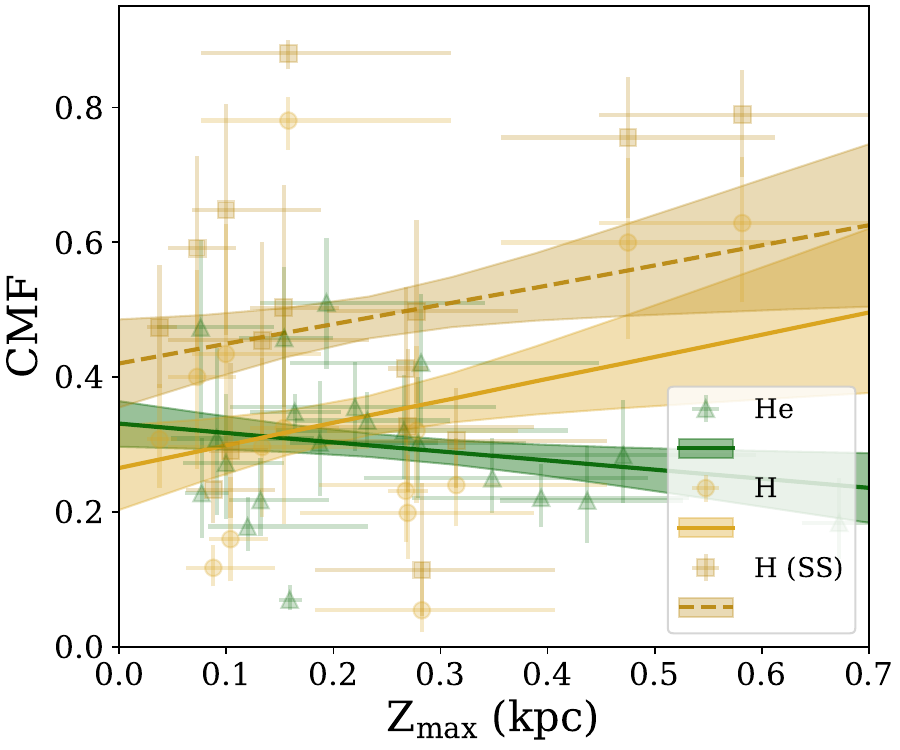}
    \includegraphics[width=1\linewidth]{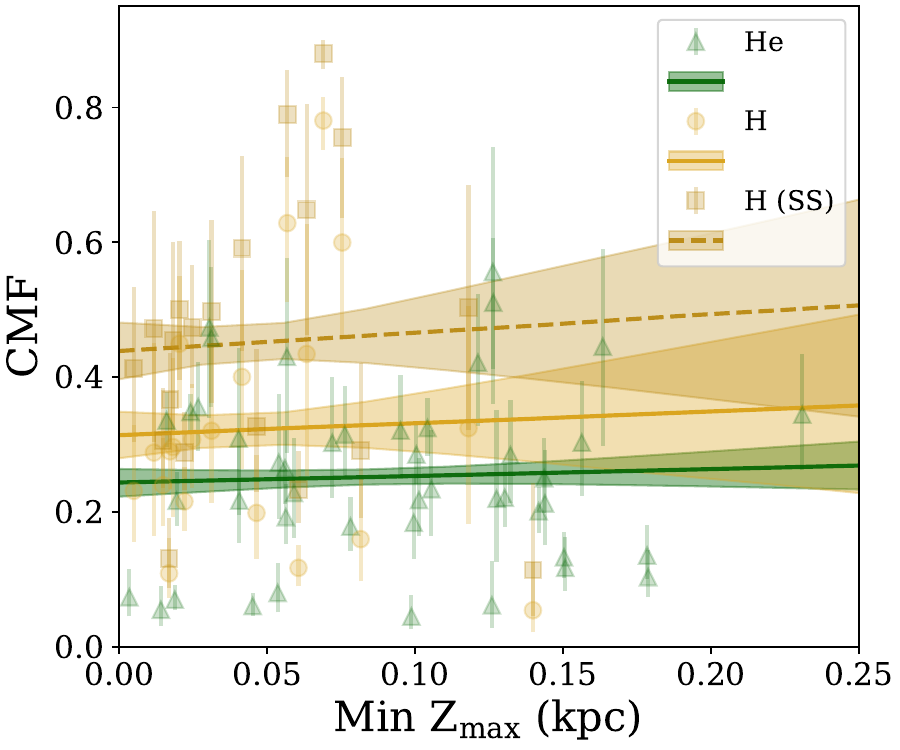}
    \caption{Calculated iron mass fraction versus relative distance from the Sun in UVW space (top), vertical amplitude (center), and minimum vertical amplitude (bottom) for the polluted WD systems, separated by type and with a steady-state adjustment applied to the H-WDs (SS). 
    Only WDs with velocity constraints reported in the literature are included in the UVW and vertical amplitude samples (Table \ref{table:WDsample}). 
    The curves show the linear fits obtained through bootstrap fitting, with 1$\sigma$ uncertainties. }
    \label{fig:cmf_kinematics_fit}
\end{figure}

In Figure \ref{fig:cmf_kinematics_fit}, we test for trends between CMF and kinematic tracers, analogous to the procedure used for total age. 
For the $R_{\rm vel}$ tracer (the distance between each star and the Sun in $V, \sqrt{U^2 + W^2}$ coordinates), we find negative slopes between CMF and $R_{\rm vel}$ for the as-observed H-, steady-state adjusted H-, and He-WDs, with probabilities of negative slope of 92\%, 89\% and 76\%, respectively. 
This is consistent with WDs with velocities closer to the high-$\alpha$/Fe region hosting material with lower CMFs. 

Considering the vertical amplitude tracer $\mathrm{Z_{max}}$, the He-WDs have a negative slope between CMF and amplitude, with a probability of $89\%$, consistent with the $R_{\rm vel}$ result. 
However, the H-WDs exhibit positive slopes between CMF and amplitude, opposite of expected if dynamical stirring increases amplitude over time. 
The probability of positive slope for the H-WDs, both as-measured and adjusted for steady state, is $90\%$. 
This positive slope hinges on the two H-WDs with high CMF and high $\mathrm{Z_{max}}$, WD0059+257 and PG0843+516; removing these two WDs results in a negative slope with a probability of $80\%$. 
While these two WDs have some of the highest $\mathrm{Z_{max}}$ values of the sample this is not necessarily consistent with their fitted ages; WD0059+257 is relatively old at about 7.5~Gyr while PG0843+516 is relatively young, at about 0.7~Gyr. 

Finally, both the H and He samples show positive slopes between CMF and the estimated minimum vertical amplitude (including all WDs, with and without reported radial velocity constraints) with probabilities of $60-70\%$. 
This trend is likely impacted by our use of minima for $\mathrm{Z_{max}}$, and may change if radial velocities can be recovered for the WDs lacking reported values.

\begin{figure}
    \includegraphics[width=1\linewidth]{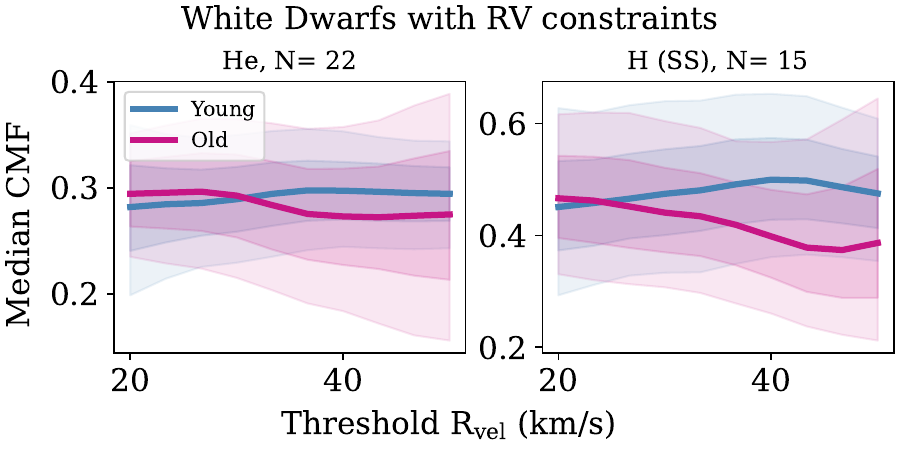}
    \includegraphics[width=1\linewidth]{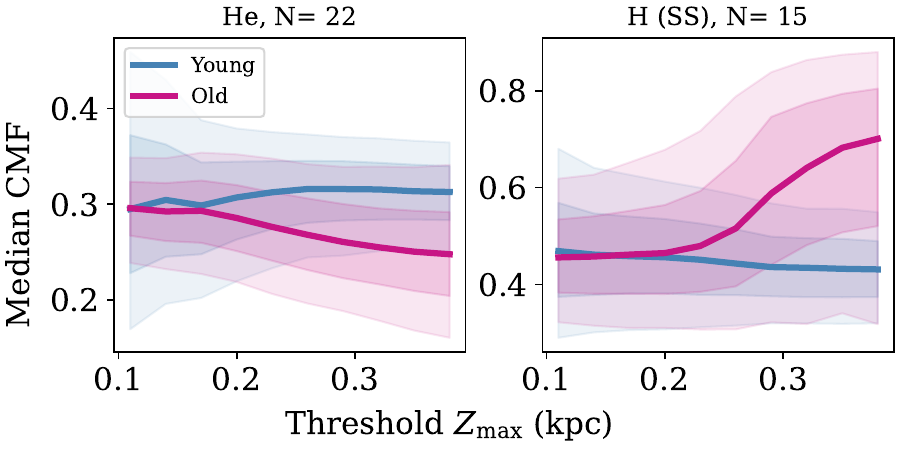}
    \includegraphics[width=1\linewidth]{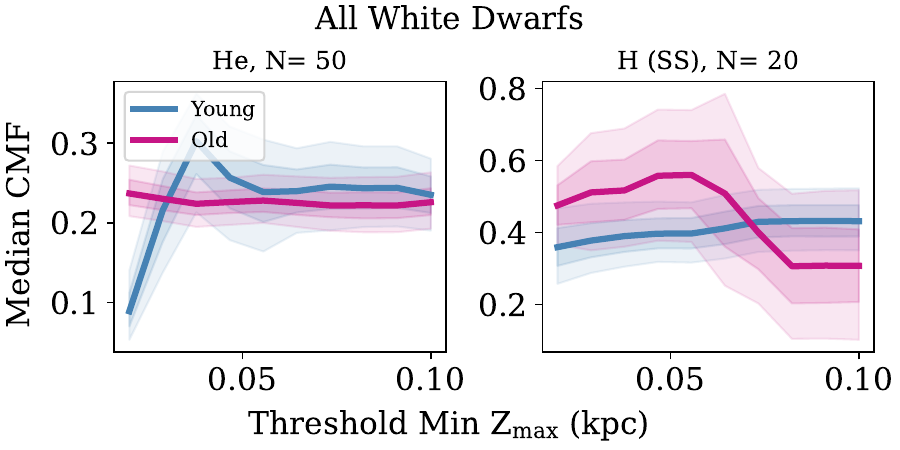}
    \caption{The median iron mass fractions (CMF) for old (pink) and young (blue) populations of WDs, where old and young are defined relative to a range of threshold values. 
    Young populations are defined as being below the reported threshold values in kinematic space. The top two rows include only the subset of WDs with radial velocity constraints, while the bottom row includes all WDs in our sample.
    Each panel corresponds with a different WD type and lists the associated sample size. 
    Abundances for H-dominated WDs are adjusted to steady-state (SS).
    The shaded regions show the $1\sigma$ and $2\sigma$ values derived from bootstrapping. }
    \label{fig:linear_med_kinematic}
\end{figure}

Figure \ref{fig:linear_med_kinematic} applies the same approach used for total age to the kinematic tracers. 
As in the slope calculations (Figure \ref{fig:cmf_kinematics_fit}), the two H-WDs WD0059+257 and PG0843+516 significantly impact the distribution of the ``Old" sample. 
For the H-WD sample we find a $<1\sigma$ separation between CMF distributions for threshold amplitudes below about 0.25~kpc, and a general reversal of the expected separation (the old population skews towards higher CMF). 
For the He-WDs, similar to the total age case, we find a $1\sigma$ separation in CMF for threshold $\mathrm{Z_{max}}$ values above about 0.2~kpc, and no significant separation at lower thresholds. 
The minimum vertical amplitude tracers show similar results, with a general reversal of the young age/high CMF trend predicted by GCE. 

\subsection{Impact of uncertainties on the significance of iron mass fraction variation at varying sample sizes}\label{section:result_popsynth}

Using the population synthesis approach described in Section \ref{section:methods_popsynth}, we estimate the number of polluted WD observations required to achieve $2-$ and $3\sigma$ separations in median CMFs for old and young systems, using the MS star distribution of CMF and age, and uncertainties drawn from the observed WD sample. 
In this case, we consider only the total age tracer, which resulted in the variation in CMF distributions most consistent with GCE predictions. 

\begin{figure}
    \centering
    \includegraphics[width=1\linewidth]{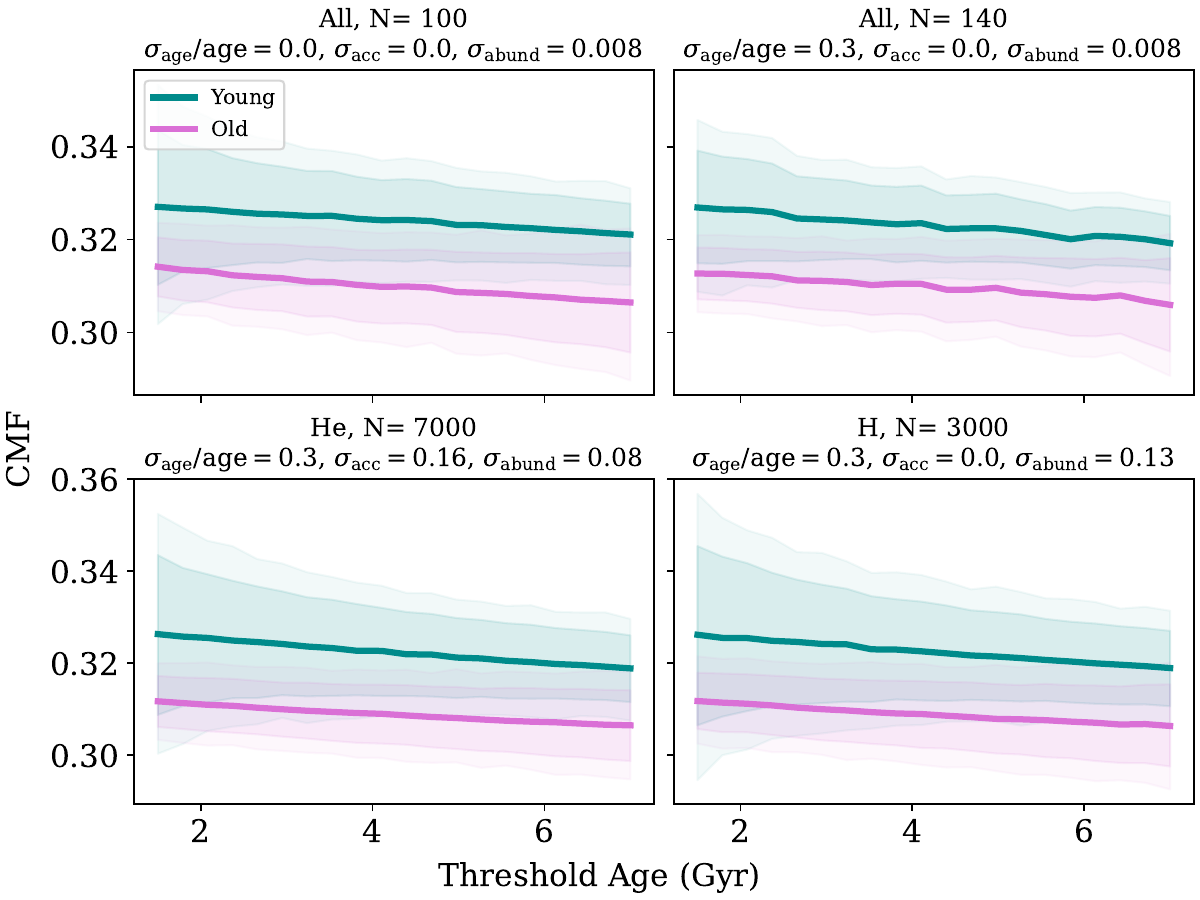}
    \caption{Simulated distribution of median iron mass fractions for old (pink) and young (green) populations of WDs, defined relative to different threshold ages (Section \ref{section:result_popsynth}). 
    The displayed sample sizes are the sizes required to achieve $3\sigma$ separations between the CMFs of the young and old systems (threshold ages of 4~Gyr and above) when incorporating different uncertainty sources. 
    The shaded regions show the $2\sigma$ and $3\sigma$ median iron mass fraction values derived from bootstrapping. 
    }
    \label{fig:popsynth_med_cmf_dist}
\end{figure}

Figure \ref{fig:popsynth_med_cmf_dist} shows the analogous results to Section \ref{section:results_cmf_wd} generated from the population synthesis.
In each panel, we show the sample size required for 3$\sigma$ separation in CMF distributions for the young and old samples, for a variety of uncertainty sources and values.
The degree of separation in CMF varies with the imposed threshold age, and we show sample sizes which achieve 3$\sigma$ significance at thresholds of 4~Gyr for consistency. 

We first impose only $\sigma_{\mathrm{abund}}=0.008$, corresponding to typical uncertainties in CMF derived from stellar abundances reported by APOGEE. 
This results in the smallest required sample size for a 3$\sigma$ separation in CMF, N=100. 
At low threshold ages, the number of stars in the ``young" group decreases, inflating uncertainties in the CMF distribution. 
Adding $\mathrm{\sigma_{age}/age}=0.3$, the typical age uncertainty for both the MS star and WD samples, moderately increases the 3$\sigma$ sample size to 140. 

Finally, we apply the uncertainty terms derived from the observed WD sample: $\mathrm{\sigma_{age}/age}=0.3$, $\mathrm{\sigma_{acc}}=$0.0, 0.16, and $\mathrm{\sigma_{abund}}=$0.13, 0.08, for the H and He WDs, respectively. 
Qualitatively, the synthetic He-WDs require roughly double the sample sizes of the H-WDs to reach 3$\sigma$ separations in CMF. 
Despite measurement uncertainties ($\mathrm{\sigma_{abund}}$) being relatively smaller for the He-WDs, their long settling timescales (resulting in increased $\mathbf{\sigma_{acc}}$) drive overall wider spreads in CMF than those of the H-WDs. 
 
Notably, the addition of all three uncertainty terms greatly inflates the predicted 3$\sigma$ sample size to several thousand WDs. 
The huge sample sizes of thousands of WDs, compared to the 20 H-WDs and 50 He-WDs in our study that result in 1-2$\sigma$ separations, are likely rooted in the simplicity of our synthesis model and uniform application of uncertainties. For this reason we focus on the relative sizes of the synthetic samples. 


\section{Discussion}\label{section:discussion}

\subsection{H- vs He-WD iron mass fractions}\label{section:HvHe_ratios}

The CMFs of the H and He WD samples differ between the values for H-WDs in steady-state and the as-measured values for H and He WDs (Figure \ref{fig:wd_cmf_dist}). 
Applying a KS test to the three samples, we find no significant difference between the H and He samples ($p=0.44$), a marginal difference between the H and H (steady-state) values ($p=0.08$), and a significant difference between the H (steady-state) and He samples ($p=0.00$). 

It is possible this H/He variation is an artifact of having a small sample; the median CMF of the three groups are within 1$\sigma$ of each other, and the shape of the steady-state distribution may originate from variations in modeling settling timescales or stellar parameters. 

Accretion phase parameters can also alter inferred CMFs. 
In this work, we use the settling grids from the MWDD to adjust the H-WD abundances to steady state. However, different grids will impact the resulting the steady-state adjusted values \citep{Buchan2025}. 
We test both the overshoot and non-overshoot models of \citealt{Koester2020} and find that the impact on steady-state CMF for the H-WDs is minimal, with variations in CMF of a few percent, within $1\sigma$ uncertainties of the as-measured values. 
Using the \citealt{Koester2020} models, the statistical significance of the H/He differences is unchanged.

It is also possible that the He-WDs are typically in decreasing accretion states, where the material is dominated by settling. 
Because Fe settles more quickly than Mg or Si, adjusting for a decreasing accretion state would increase the CMFs of the material polluting He-WDs. 
However, calculated CMF is very sensitive to the assumed time in the decreasing phase, increasing by factors of a few within a few settling timescales, and constraining the observed time accretion phase is generally uncertain \citep[e.g.,][]{Trierweiler2025}. 
As a test, we apply a uniform adjustments assuming we observe the He-WDs at a time of about 5 settling timescales after steady state, with an accretion disk timescale of $10^{5.6}~\mathrm{yr}$, but find that the adjusted He-WDs are still statistically distinct from the steady-state H-WDs, with a p-value of 0.02. 
Individually tuning the He-WD accretion phases likely can bring the two populations into agreement, particularly if more extreme accretion disk timescales are allowed, however we do not attempt to do so in this work. 

Observational uncertainty may also impact abundances. 
There is a noted discrepancy between Si and Fe abundances determined from optical versus UV spectra which may impact silicate masses \citep[e.g.,][]{Gansicke2012, Jura2012, Xu2019, Williams2025}. 
For all elements used in this work we take an average of the optical and UV results when both are reported, unless the corresponding reference has recommended the use of a particular value.
Additionally, while we typically draw abundances from a single study of each WD, as reported in Table \ref{table:WDsample}, variations in abundance measurements can arise between different studies of the same object. 
For example, for Ton345 we find a CMF of about $43\%$ using abundances from \citealt{Wilson2015} and a CMF of about $30\%$ using abundances from \citealt{Jura2015}, though the two results are within $1\sigma$ of each other. 
Such observational uncertainties could therefore introduce additional scatter in the CMFs presented in this work. 

Alternatively, the difference could be influenced by physics in the WD atmospheres. 
On the hot end of the H-WD sample ($\gtrsim15000$~K) radiative levitation may affect abundances of Si \citep{Chayer2010}. 
Based on the models of \citealt{Koester2014}, this should not be a dominant effect as all of the H-WDs in our sample have $\mathrm{\log (Si/H)}$  abundances greater than the equilibrium values that can result from radiative levitation, and the impact on CMF should be very minimal as the levitation should not affect Mg or Fe abundances. 
However, radiative levitation should be considered as observed H-WD samples continue to grow, especially on the high temperature end. 

Finally, all H-WDs in our sample have $\mathrm{T_{eff}>10,000~K}$, falling in the regime where thermohaline mixing may dominate over gravitational settling \citep{Bauer2018, Bauer2019, Buchan2025,Deal2026}. 
In this scenario, the additional mixing causes observed element ratios to be a direct match to the parent body, in which case the `true' distribution of CMFs for the H-WDs should match the as-measured H-WD case, not the steady state scenario. 
Invoking thermohaline would thus bring the H and He CMF distributions into agreement, and imply that the non-adjusted values listed throughout this work are most accurate for the H sample. 

\subsection{Trends in trace core elements}\label{section:trace_elements}

\begin{figure}
    \centering
    \includegraphics[width=\linewidth]{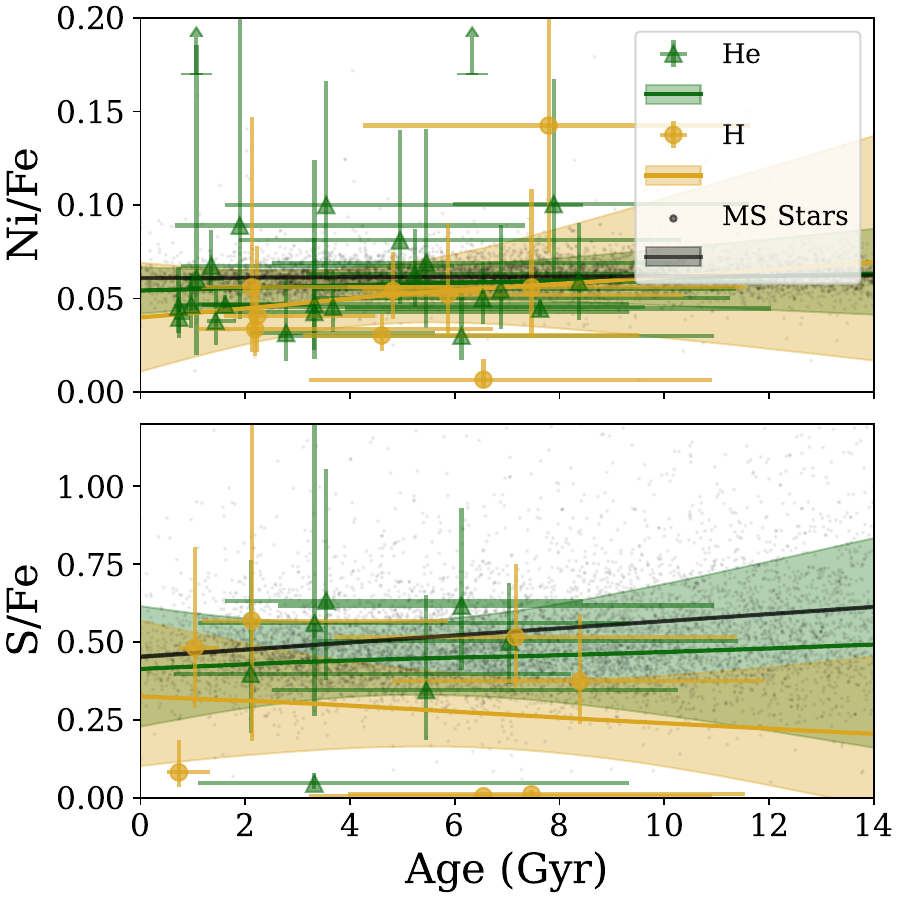}
    \caption{Number ratios of core-forming elements (Ni/Fe, and S/Fe) versus stellar age, for the WDs (colored points) and MS stars (black points). H-WD values have been adjusted to steady state. The shaded regions show linear fits to each sample, with $1\sigma$ uncertainties. }
    \label{fig:core_element_trends}
\end{figure}

Throughout this work we focus on iron as the tracer of metal core material. 
In this section, we consider additional trace elements which can also partition into core material, Ni and S. 
While Ni can function as an identifier for core-rich material, S may be useful in roughly constraining formation conditions of the accreted material.
In Figure \ref{fig:core_element_trends}, we show the number ratios of Ni and S relative to Fe, as a function of total calculated age, for the WD and MS star samples. 
Ni and S abundance measurements for the WDs are more rare than the major rock elements; within our sample, 34 and 15 WDs have Ni and S reported, respectively. 

Ni/Fe ratios for the polluted WDs are largely near bulk Earth and CI chondrite values (about 0.05, by number), though with a large amount of scatter.
As shown in the MS star sample, Ni/Fe ratios do not change significantly with GCE. 
The scatter in Ni/Fe ratios should thus reflect the effects of core-mantle sampling and the oxidation of the parent body. 
Two He WDs, WD1244+498 and 0930+0618, have extremely high Ni/Fe ratios, potentially indicating Ni-rich cores within more oxidized bodies. 
Neglecting these WDs, Ni/Fe ratios follow a very faint positive slope with age (probabilities of about 60\% for both He and H WDs in both cases), roughly consistent with the GCE trend. 

In general, S/Fe number ratios amongst WD pollution tend to be much higher than that of bulk Earth (S/Fe$\approx0.03$), instead trending towards CI chondrites (S/Fe$\approx0.5$). 
This could be interpreted as a signature of pollution originating in the outer solar system, consistent with the relatively high masses and O-excesses observed for many polluted WDs. 
As S is an $\alpha$-element, S/Fe ratios among MS stars generally increase with age. 
We see this increase reflected in the He-WD sample, but not in the H-WD sample, though the sample sizes for both populations are very small.
Further analyses of the systematics of Ni and S detections in conjunction with Fe measurements may be useful in constraining the relative compositions and formation conditions of of the debris. 

\subsection{Increasing polluted WD sample sizes and improving age constraints}
In Section \ref{section:result_popsynth}, we estimate on the order of a thousand observed polluted WD systems, spanning a full range of stellar ages, are required to achieve a $3\sigma$ separation in median CMFs between young and old systems, for both H and He WDs. 
This comprises a roughly 100 fold increase from the current observed sample; while we point out in Section \ref{section:result_popsynth} that the estimated sample size is likely an overestimate, we now discuss prospects for increasing the observed polluted WD system. 

The current sample of observed polluted WDs, often with combined observations in the optical and UV from instruments such as HIRES (Keck I Telescope at Maunakea Observatory, \citealt{Vogt1994}), MIKE (Clay Telescope at Las Campanas Observatory, \citealt{Bernstein2003}), and COS (Hubble Space Telescope, \citealt{green12}), reaches magnitudes of $G\sim16-17$. 
Considering the sample of WDs identified from \textit{Gaia} photometry by \cite{GentileFusillo2021a}, sensitive to WDs of magnitude $G < 20-21$ \citep{GaiaCollaboration2016}, about 2500 fall within the current limits for high-resolution follow-up. 
Therefore, as a lower limit, about a few percent of the Gaia-identified WD census make up the current sample of observed highly-polluted WDs. 

Next-generation high-resolution spectrographs such as the Extremely Large Telescope's ANDES (ArmazoNes high Dispersion Echelle Spectrograph) are anticipated to push the limiting magnitudes for WD pollution characterization to about $G\sim 19$ \citep{Maiolino2013}. 
Returning to the \cite{GentileFusillo2021a} catalog, this increases the number of WDs within the range of high-resolution optical spectroscopy by an order of magnitude, a promising opportunity for more sensitive measurements of the impact of galactic evolution of exoplanetary material. 
Additionally, spectroscopic surveys like DESI \citep{DESICollaboration2026} have already yielded new polluted WD abundance characterizations \citep{Izquierdo2026}, and current and future surveys such as 4MOST \citep{deJong2019}, WEAVE \citep{Dalton2012, Jin2024}, SDSS-V Milky Way Mapper \citep{Almeida2023, Meszaros2025}, and the future HWO \citep{Feinberg2024} are also poised to rapidly increase the sample of identified WDs.

In addition to abundance measurements, tests of galactic physics on pollution compositions require reliable age and kinematic constraints. 
It would therefore be valuable to increase the sample of observed polluted WDs (with several measured elements) in wide binaries with MS companions, in order to more tightly constrain the MS lifetime component of total age from the companion star \citep[e.g.,][]{Rebassa-Mansergas2025}. 
Such samples also provide a chance to compare debris compositions to host star abundances via the MS companion \citep{Bonsor2021, Aguilera-Gomez2025}, helping to constrain the amount of inherent composition scatter between star and planet material due to planet formation.
The frequency of pollution on WDs in binaries appears to be lower than that of single WDs \citep{O'Brien2024, Xu2026}, so a polluted WD/MS sample may be subject to some biases compared to the single WD population but would still be valuable in probing variation across age. 

Finally, as the sample of WDs with velocity constraints grows, we can better leverage age-velocity dispersion relations to constrain the galactic populations for polluted WDs \citep[e.g.,][]{Raddi2022}. 
In our sample, 35 WDs have reported RVs. 
We can broadly check the velocity dispersions of old and young stars by splitting this RV sample at arbitrary age thresholds, as reported by the \textsc{wdwarfdate} fits. 
In general, for threshold ages around $3-5~\rm Gyr$, the older sample tends to have larger velocity dispersions in U, V, and W, however the small size of the sample and large age uncertainties prevent more detailed analysis. 
In addition to providing more specific galactic context for polluted systems, an expanded RV sample would help assess the more massive WDs and identify systems that might have anomalous ages due to forming from double-WD mergers \citep{Cheng2020}.  

\section{Conclusions}
In this work we test the impact of galactic chemical evolution (GCE) on exoplanetary rock compositions using polluted WDs. 
GCE is expected to influence exoplanet iron mass fraction/core mass fractions (CMFs) because it shapes long-term, metallicity-dependent variations in the relative abundances of alpha-elements (or equivalently, lithophile elements) to Fe. 
We investigate this relation by calculating CMFs from measured element abundances for WD pollutants, and using total and kinematic ages for WDs as tracers of their formation compositions. 
Throughout this work, we benchmark results from WD pollution against those for MS stars. 

Within our sample of polluted WDs, we generally find the expected negative trend between CMF and total age, though age estimates typically have very large uncertainties. 
This negative trend with CMF is stronger for WDs with atmospheres dominated by H than for those dominated by He. 
For WDs with reported radial velocity constraints, comparisons between CMF and galactic velocities are also consistent with younger systems having debris with relatively larger CMFs. 
However, this result reverses when using vertical orbital amplitude through the galactic disk as a kinematic tracer instead. 
Using minimum estimates for vertical orbital amplitude for WDs without reported radial velocities, we find faint but positive trends between CMF and amplitude. 


Finally, using a simple population synthesis model based on observed polluted WDs, we show the dependencies between separations in CMF for old and young populations of stars and different sources of uncertainty in age and CMF. 
We show that He-WDs require larger sample sizes, due to increased uncertainty in the phase of accretion of the WD, which leads to a larger range of potential CMF values for the accreted body. 
This may also explain why trends between CMF and age are much stronger amongst the H-WD sample as compared to the He-WDs. 
As the sample of observed polluted WDs increases, continuing to combine abundances with stellar parameters will help untangle potential large-scale variations in exoplanetary material from composition trends due to accretion physics and planet-formation driven variations in composition between individual systems. 

\begin{acknowledgments}
We thank Carl Melis, J. J. Hermes, and Catherine Manea for helpful discussions and feedback on this manuscript. 
We also thank the Rice Research Group and the HST 17822 collaboration for helpful conversations over the course of this work.
This work was supported by Heising-Simons Foundation Grants \#2021-2802 and \#2023-4478, as well as National Geographic Grant EC-115062R-24.
The research shown here acknowledges use of the Hypatia Catalog, an online compilation of stellar abundance data as described in \citet{Hinkel2014}.
\end{acknowledgments}

\software{astropy \citep{astropy:2013, astropy:2018, astropy:2022}, wdwarfdate \citep{Kiman2022}, Gala\citep{Price-Whelan2017}, NumPy \citep{Harris2020}, SciPy \citep{Virtanen2020}, Matplotlib \citep{Hunter2007}}

\appendix
\section{Updated WD Parameters}

Table \ref{table:MWDD_recent_WDsample} lists the WD parameters listed in the MWDD for each WD in our sample, with the corresponding reference and total age fitted with \textsc{wdwarfdate}. 

\startlongtable
\tabletypesize{\scriptsize}
\centerwidetable
\begin{deluxetable*}{l|l|l|l|l|l}
\tablecaption{WD sample, using the WD parameters most recently reported by the MWDD, with the corresponding reference listed in the table. The order of WDs is the same as in Table \ref{table:WDsample}.}\label{table:MWDD_recent_WDsample}
\tablehead{\colhead{Name in Reference} & \colhead{Type} & \colhead{$\mathrm{T_{eff}}$ (K)} & \colhead{log $g$ $\mathrm{(cm/s^2)}$} & \colhead{Age\hyperlink{a}{$^a$}} & \colhead{Reference\hyperlink{b}{$^b$}}}
\startdata
NLTT\,43806 & H & 5951$\pm$18 & 8.25$\pm${0.01} & 4.9$\pm^{0.2}_{0.1}$ & \citealt{Kilic2026b} \\
LSPM\,J0207+3331 & H & 5910$\pm$98 & 8.11$\pm${0.03} & 4.8$\pm^{1.7}_{0.6}$ & \citealt{LeBourdais2025} \\
WD\,2157-574 & H & 7114$\pm$55 & 8.10$\pm${0.02} & 3.3$\pm^{1.7}_{0.5}$ & \citealt{O'Brien2024} \\
WD\,2115-560 & H & 9529$\pm$76 & 7.97$\pm${0.02} & 6.9$\pm^{4.5}_{3.2}$ & \citealt{O'Brien2024} \\
G29$-$38 & H & 11529$\pm$206 & 8.04$\pm${0.03} & 3.2$\pm^{3.9}_{1.3}$ & \citealt{O'Brien2024} \\
WD\,0145+234 & H & 12680$\pm$80 & 8.06$\pm${0.01} & 2.2$\pm^{2.2}_{0.6}$ & \citealt{Trierweiler2025} \\
GD\,56 & H & 14919$\pm$152 & 8.01$\pm${0.01} & 3.5$\pm^{4.0}_{1.6}$ & \citealt{Kilic2026a} \\
WD\,0842+572 & H & 15400$\pm$70 & 7.91$\pm${0.01} & 7.8$\pm^{3.8}_{3.5}$ & \citealt{Trierweiler2025} \\
J0611$-$6931 & H & 17284$\pm$335 & 7.95$\pm${0.02} & 5.7$\pm^{4.6}_{2.9}$ & \citealt{Vincent2024} \\
HE\,0106$-$3253 & H & 17350$\pm$200 & 8.12$\pm${0.10} & 1.6$\pm^{3.6}_{0.8}$ & \citealt{Xu2019} \\
SDSS\,J1043+0855 & H & 15942$\pm$0 & 7.93$\pm${0.00} & 6.2$\pm^{4.6}_{3.3}$ & MWDD H \\
PG\,1015+161 & H & 19228$\pm$346 & 8.03$\pm${0.01} & 2.1$\pm^{2.5}_{0.7}$ & \citealt{Kilic2025} \\
WD\,1943+163 & H & 17271$\pm$0 & 7.80$\pm${0.00} &   & MWDD H \\
WD\,1953–715 & H & 18975$\pm$258 & 7.96$\pm${0.09} & 4.2$\pm^{5.2}_{2.6}$ & \citealt{Koester2014} \\
WD\,0059+257 & H & 20491$\pm$278 & 8.00$\pm${0.09} & 3.1$\pm^{5.1}_{1.7}$ & \citealt{Koester2014} \\
SDSS\,1228+1040 & H & 23500$\pm$200 & 8.16$\pm${0.10} & 0.9$\pm^{1.3}_{0.4}$ & \citealt{Xu2019} \\
GALEX\,1931+0117 & H & 20409$\pm$0 & 7.95$\pm${0.00} & 4.7$\pm^{5.0}_{2.6}$ & MWDD H \\
J2100+2122 & H & 21997$\pm$0 & 7.92$\pm${0.00} & 5.4$\pm^{4.8}_{3.0}$ & MWDD H \\
WD\,1647+375 & H & 22938$\pm$222 & 7.98$\pm${0.01} & 3.2$\pm^{3.9}_{1.6}$ & \citealt{Kilic2026a} \\
PG\,0843+516 & H & 24670$\pm$400 & 7.93$\pm${0.10} & 3.8$\pm^{5.4}_{2.4}$ & \citealt{Xu2019} \\
\hline
SDSS\,J1535+1247 & He & 5897$\pm$14 & 8.16$\pm${0.01} & 5.4$\pm^{0.6}_{0.4}$ & \citealt{Kilic2026b} \\
SDSS\,J1038$-$0036 & He & 7865$\pm$24 & 8.20$\pm${0.01} & 2.4$\pm^{0.4}_{0.2}$ & \citealt{Kilic2026b} \\
J0956 & He & 8712$\pm$66 & 8.18$\pm${0.02} & 2.2$\pm^{0.5}_{0.3}$ & \citealt{Kilic2026b} \\
1252+7352 & He & 8574$\pm$134 & 8.19$\pm${0.03} & 2.2$\pm^{0.5}_{0.3}$ & \citealt{Kilic2026b} \\
1352+0323 & He & 9612$\pm$131 & 8.07$\pm${0.07} & 4.1$\pm^{5.1}_{1.9}$ & \citealt{Kilic2026b} \\
WD\,0446$-$255 & He & 9934$\pm$108 & 7.85$\pm${0.02} &   & \citealt{Vincent2024} \\
J2050 & He & 9805$\pm$266 & 8.17$\pm${0.12} & 2.6$\pm^{4.7}_{1.0}$ & \citealt{Kilic2026b} \\
0452$-$0214 & He & 9677$\pm$114 & 7.92$\pm${0.02} & 9.6$\pm^{3.2}_{3.6}$ & \citealt{Kilic2026b} \\
GD\,362 & He & 9928$\pm$78 & 8.02$\pm${0.01} & 6.1$\pm^{4.6}_{2.8}$ & \citealt{Kilic2026b} \\
PG\,1225$-$079 & He & 11248$\pm$126 & 8.18$\pm${0.02} & 1.5$\pm^{0.5}_{0.3}$ & \citealt{O'Brien2024} \\
WD\,1350$-$162 & He & 12480$\pm$252 & 8.09$\pm${0.03} & 2.5$\pm^{3.1}_{0.8}$ & \citealt{Vincent2024} \\
WD\,1232+563 & He & 11499$\pm$266 & 8.04$\pm${0.03} & 4.6$\pm^{4.7}_{2.3}$ & \citealt{Kilic2026a} \\
1626+3136 & He & 11832$\pm$248 & 8.16$\pm${0.07} & 1.9$\pm^{2.5}_{0.6}$ & \citealt{Kilic2026a} \\
0242+0426 & He & 11760$\pm$182 & 7.95$\pm${0.06} & 7.1$\pm^{4.5}_{3.6}$ & \citealt{Kilic2026a} \\
0850+3208 & He & 11799$\pm$150 & 8.13$\pm${0.02} & 1.9$\pm^{1.1}_{0.5}$ & \citealt{Kilic2026a} \\
1333+3254 & He & 12245$\pm$402 & 8.00$\pm${0.08} & 5.3$\pm^{5.2}_{3.0}$ & \citealt{Kilic2026a} \\
SDSS\,J2339$-$0424 & He & 13322$\pm$171 & 7.91$\pm${0.02} & 9.1$\pm^{3.3}_{3.6}$ & \citealt{Kilic2026a} \\
SDSS\,J0738+1835 & He & 14182$\pm$247 & 8.06$\pm${0.03} & 3.2$\pm^{4.1}_{1.5}$ & \citealt{Kilic2026b} \\
HS\,2253+8023 & He & 14000$\pm$0 & 8.10$\pm${0.00} & 2.3$\pm^{3.5}_{0.9}$ & \citealt{Klein2011} \\
WD\,1425+540 & He & 14093$\pm$406 & 8.00$\pm${0.05} & 5.5$\pm^{4.9}_{3.0}$ & \citealt{Kilic2026a} \\
0944$-$0039 & He & 12321$\pm$180 & 8.10$\pm${0.03} & 2.3$\pm^{2.7}_{0.7}$ & \citealt{Kilic2026a} \\
Gaia\,J0218+3625 & He & 14700$\pm$500 & 7.86$\pm${0.05} &   & \citealt{Doyle2023} \\
EC\,22211$-$2525 & He & 14740$\pm$500 & 7.89$\pm${0.05} & 8.4$\pm^{3.7}_{3.8}$ & \citealt{Doyle2023} \\
WD\,2207+121 & He & 14752$\pm$1192 & 7.97$\pm${0.12} & 5.0$\pm^{5.3}_{3.1}$ & \citealt{Xu2019} \\
WD\,1551+175 & He & 15011$\pm$182 & 7.98$\pm${0.02} & 6.6$\pm^{4.4}_{3.3}$ & \citealt{Kilic2026a} \\
WD\,1244+498 & He & 15150$\pm$500 & 7.97$\pm${0.05} & 6.3$\pm^{4.7}_{3.4}$ & \citealt{Doyle2023} \\
WD\,1248+1004 & He & 15180$\pm$500 & 8.11$\pm${0.05} & 2.0$\pm^{3.0}_{0.7}$ & \citealt{Doyle2023} \\
G241$-$6 & He & 13947$\pm$0 & 7.95$\pm${0.00} & 7.1$\pm^{4.4}_{3.6}$ & MWDD He \\
GD\,40 & He & 13249$\pm$88 & 7.95$\pm${0.01} & 8.2$\pm^{3.8}_{3.6}$ & \citealt{Kilic2026a} \\
1516$-$0040 & He & 13240$\pm$172 & 7.93$\pm${0.02} & 8.6$\pm^{3.6}_{3.6}$ & \citealt{Kilic2026a} \\
Gaia\,J1922+4709 & He & 15500$\pm$500 & 7.95$\pm${0.05} & 6.9$\pm^{4.4}_{3.6}$ & \citealt{Doyle2023} \\
WD\,1145+017 & He & 14434$\pm$250 & 8.06$\pm${0.02} & 3.1$\pm^{3.8}_{1.4}$ & \citealt{Kilic2026a} \\
GD\,378 & He & 15453$\pm$128 & 7.92$\pm${0.01} & 8.7$\pm^{3.5}_{3.6}$ & \citealt{Kilic2025} \\
0859+1123 & He & 15928$\pm$715 & 8.17$\pm${0.09} & 1.6$\pm^{2.9}_{0.7}$ & \citealt{Kilic2026a} \\
0030+1526 & He & 15160$\pm$288 & 8.07$\pm${0.03} & 2.7$\pm^{3.7}_{1.1}$ & \citealt{Kilic2026a} \\
0930+0618 & He & 15901$\pm$442 & 8.06$\pm${0.04} & 3.0$\pm^{4.3}_{1.5}$ & \citealt{Kilic2026a} \\
1627+1723 & He & 16100$\pm$623 & 8.14$\pm${0.06} & 1.6$\pm^{2.2}_{0.6}$ & \citealt{Kilic2026a} \\
1109+1318 & He & 15626$\pm$496 & 8.09$\pm${0.08} & 2.6$\pm^{4.8}_{1.3}$ & \citealt{Kilic2026a} \\
SDSS\,J1734+6052 & He & 16403$\pm$341 & 8.00$\pm${0.02} & 5.5$\pm^{4.7}_{2.8}$ & \citealt{Kilic2026a} \\
0259$-$0721 & He & 13524$\pm$382 & 7.96$\pm${0.04} & 7.1$\pm^{4.4}_{3.5}$ & \citealt{Kilic2026a} \\
GD\,424 & He & 17201$\pm$440 & 8.30$\pm${0.02} & 0.6$\pm^{0.1}_{0.1}$ & \citealt{Kilic2025} \\
1359$-$0217 & He & 13868$\pm$176 & 7.72$\pm${0.04} &   & \citealt{Kilic2026a} \\
J0644$-$0352 & He & 17800$\pm$532 & 8.04$\pm${0.03} & 3.5$\pm^{4.4}_{1.8}$ & \citealt{Vincent2024} \\
GD\,61 & He & 15680$\pm$1633 & 8.04$\pm${0.00} & 4.0$\pm^{4.9}_{2.1}$ & \citealt{Coutu2019} \\
WD\,1415+234 & He & 17300$\pm$500 & 8.17$\pm${0.05} & 1.2$\pm^{0.8}_{0.4}$ & \citealt{Doyle2023} \\
SDSS\,J2248+2632 & He & 17323$\pm$291 & 7.98$\pm${0.02} & 6.2$\pm^{4.5}_{3.2}$ & \citealt{Kilic2026a} \\
WD\,J2047$-$1259 & He & 17105$\pm$0 & 8.06$\pm${0.00} & 3.0$\pm^{4.5}_{1.5}$ & MWDDD He \\
Ton\,345 & He & 19432$\pm$0 & 8.15$\pm${0.00} & 1.3$\pm^{1.1}_{0.5}$ & \citealt{Genest-Beaulieu2019} \\
WD\,1536+520 & He & 20199$\pm$0 & 7.87$\pm${0.00} & 8.2$\pm^{3.7}_{3.7}$ & \citealt{Genest-Beaulieu2019} \\
WD1622+587 & He & 18848$\pm$0 & 7.85$\pm${0.00} &   & MWDD He \\
\enddata
\hypertarget{a}{a}: Age calculated as the sum of the cooling age of the WD and the main sequence lifetime (Section \ref{section:ages})\\
\hypertarget{b}{b}: ``MWDD" references indicate parameters directly reported in the MWDD, with the corresponding model indicated. )
\end{deluxetable*}

\bibliography{bibli}{}
\bibliographystyle{aasjournalv7}

\end{document}